\documentclass[reprint, amsmath, pra,superscriptaddress]{revtex4-2}

\usepackage{graphicx}% Include figure files
\usepackage{epsfig}
\usepackage{dcolumn}% Align table columns on decimal point
\usepackage{bm}% bold math
\usepackage{blkarray} % blockarray for the Hamiltonian matrix

\newcommand{\cmm}{\ensuremath{\mathrm{cm}^{-1}}}
\newcommand{\Xstate}{$X^1\Sigma^+$ }
\newcommand{\Bstate}{$B^1\Pi$ }
\newcommand{\astate}{$a^3\Sigma^+$ }
\newcommand{\cstate}{$c^3\Sigma^+$ }
\newcommand{\Estate}{$E$(4)$^1\Sigma^+$ }
\newcommand{\comment}[1]{}
\graphicspath{ {./images/} }

\begin{document}

\title{The \astate state of KCs revisited: hyperfine structure analysis and potential refinement}

\author{V. Krumins}
\author{M. Tamanis}
\author{R. Ferber}
\affiliation{Laser Center, Faculty of Physics, Mathematics and Optometry, University of Latvia, 19 Rainis blvd, Riga LV-1586, Latvia}

\author{A. V. Oleynichenko}
\affiliation{B. P. Konstantinov Petersburg Nuclear Physics Institute of National Research Centre “Kurchatov Institute”, Gatchina, 188300 Leningrad District, Russia}

\author{L. V. Skripnikov}
\affiliation{B. P. Konstantinov Petersburg Nuclear Physics Institute of National Research Centre “Kurchatov Institute”, Gatchina, 188300 Leningrad District, Russia}
\affiliation{Saint Petersburg State University, 7/9 Universitetskaya nab., St. Petersburg 199034 Russia}

\author{A. Zaitsevskii}
\affiliation{B. P. Konstantinov Petersburg Nuclear Physics Institute of National Research Centre “Kurchatov Institute”, Gatchina, 188300 Leningrad District, Russia}
\affiliation{Department of Chemistry, Lomonosov Moscow State University, 119991, Moscow, Leninskie gory 1/3, Russia}

\author {E. A. Pazyuk}
\author{A. V. Stolyarov}
\affiliation{Department of Chemistry, Lomonosov Moscow State University, 119991, Moscow, Leninskie gory 1/3, Russia}

\author{A. Pashov}
\email{pashov@phys.uni-sofia.bg}
\affiliation{Faculty of Physics, Sofia University, 5 James Bourchier Boulevard, 1164 Sofia, Bulgaria}

\date{\today}

\begin{abstract}
Laser-induced fluorescence spectra of the $c^3\Sigma^+(v_{c},J_{c}=N_{c})\rightarrow a^3\Sigma^+(v_{a},N_{a} = J_{c} \pm 1)$ transitions excited from the ground $X^1\Sigma^+$ state of $^{39}$K$^{133}$Cs molecule were recorded with Fourier-transform spectrometer IFS125-HR (Bruker) at the highest achievable spectral resolution of 0.0063 \cmm. Systematic study of the hyperfine structure (HFS) of the \astate state for levels with $v_{a} \in [0, 27]$ and $N_{a} \in [24, 90]$ shows that the splitting monotonically increases with $v_{a}$. The spectroscopic study was supported by \emph{ab initio} calculations of the magnetic hyperfine interaction in \Xstate and \astate states. The discovered variation of the electronic matrix elements with the internuclear distance $R$ is in a good agreement with the observed  $v_{a}$-dependencies of the HFS. Overall set of available experimental data on the \astate state was used to improve the potential energy curve particularly near a bottom, providing the refined dissociation energy $D_e$=267.21(1)~\cmm. The \emph{ab initio} HFS matrix elements, combined with the empirical \Xstate and \astate PECs in the framework of the invented coupled-channel deperturbation model, reproduce the experimental term values of both ground states within 0.003~\cmm\ accuracy up to their common dissociation limit.
\end{abstract}
\maketitle

\section{Introduction}

The KCs molecule is the last member of the group of alkali-metal diatomics, which was studied by high-resolution spectroscopy only in 2008 \cite{Ferber:2008} followed by more detailed studies in 2009 \cite{Ferber:2009}. Since then tens of research papers have been published highlighting the growing interest of the scientific community to this molecule. Among these one can see purely spectroscopic studies, \emph{ab initio} electronic structure calculations, laser manipulation of cold and ultracold atoms and advanced deperturbation analysis going beyond the conventional adiabatic (Born-Oppenheimer) approximation~\cite{Pazyuk:2019}. Such an interest to this molecule may be explained by the prospects of variety of experiments with cold KCs molecules facilitated by convenient optical transitions, which give access to various electronic states due to strong spin-orbit interactions.

The KCs molecule attracted interest of theoreticians somewhat earlier; in 2000 Korek et al. reported the first \emph{ab initio} potential energy curves (PECs) \cite{Korek:2000}. Later, in Ref.~\cite{Korek:2006,Kim:2009} new PECs and various matrix non-adiabatic elements were calculated, necessary to treat the strongly coupled excited electronic states. The series of theoretical papers on KCs was supplemented by models of the hyperfine structure (HFS) of excited electronic states \cite{Orban:2019}.

After first experimental studies of the ground state at high resolution \cite{Ferber:2008,Ferber:2009} several groups undertook systematic experiment based studies of excited states. The group of states correlated to the first excited atomic asymptote K($4^2S$)+Cs($6^2P$) was studied in detail, namely for $A^1\Sigma^+$ \cite{Kruzins:2010,Kruzins:2013}, $b^3\Pi$ \cite{Tamanis:2010}, $B^1\Pi$ \cite{Birzniece:2015}, and $c^3\Sigma^+$ \cite{Szczepkowski:2018,Kruzins:2021} states. Several higher electronic states were also extensively studied:
$D$(2)$^1\Pi$ \cite{Birzniece:2015b,Szczepkowski:2020a},
(4)$^1\Pi$ \cite{Szczepkowski:2013},
(5)$^1\Pi$ \cite{Szczepkowski:2015},
$C$(3)$^1\Sigma^+$ \cite{Szczepkowski:2020b},
(4)$^1\Sigma^+$ \cite{Busevica:2011,Szczepkowski:2012}, and
(6)$^1\Sigma^+$ \cite{Szczepkowski:2014,Szczepkowski:2019}.
A number of excited states studies were carried out by advanced multi-channel deperturbation models, which managed to reach the experimental accuracy of about 0.01 \cmm~even for heavily spin-orbit coupled systems \cite{Tamanis:2010,Kruzins:2013,Szczepkowski:2019,Szczepkowski:2020a}.

Prospects for formation of cold KCs molecules have been evaluated in Refs.~\cite{Klincare:2012, Borsalino:2016}, based on the analysis of the ground and several excited states of the molecule. Experiments with mixed ultracold ensembles of K and Cs were reported in 2014 - 2017~\cite{Patel:2014, Groebner:2016, Groebner:2017}.

A comprehensive study of the lowest triplet \astate state, along with the ground  \Xstate state was performed in \cite{Ferber:2009,Ferber:2013} by Fourier-transform-spectroscopy (FTS) of laser-induced-fluorescence (LIF). Since it is a triplet state, interaction between the electron spin and the nuclear spin needed to be taken into account. It was done, as for $^{23}$Na$^{85}$Rb in Ref.~\cite{Kasahara:1996}, by assuming the Fermi contact (FC) interaction as the leading term responsible for the HFS of the \astate state. In a later paper \cite{Ferber:2013}, new experimental data led to an improved description of the levels near the K($4^2S$)+Cs($6^2S$) atomic asymptote. In this region it is important to take into account the HFS interaction between the \astate and the \Xstate states, and this was done in Refs.~\cite{Ferber:2009,Ferber:2013} by building a coupled channels model. In both studies the electronic matrix elements of the FC interaction with the K nucleus, $A^{\mathrm{K}}$ and the Cs nucleus, $A^{\mathrm{Cs}}$ were fixed to the asymptotic values, corresponding to the atomic HFS constants. It was found that this is sufficient to describe both the splitting of the \astate levels and the level shifts due to \astate-\Xstate interaction within the experimental uncertainty.

Later, the \astate state was a subject of several studies. In Ref.~\cite{Groebner:2017} the repulsive part of the potential curves from~\cite{Ferber:2013} was further adjusted in order to reproduce the positions of newly observed Feshbach resonances. Very recently Schwarzer and Toennies~\cite{Schwarzer:2021} reanalyzed the experimental data from~\cite{Ferber:2009} using the semi-empirical analytical form of the PEC, which was different from the PEC form used in Refs.~\cite{Ferber:2009,Ferber:2013}. Since the conventional adiabatic approach was applied, the PEC from~\cite{Schwarzer:2021} is able to reproduce experimental data only up to $v_{a}\le20$ since above $v_{a} = 20$ the mixing with the \Xstate state becomes important.

In 2020 Oleynichenko et al. \cite{Oleynichenko:2020} applied relativistic calculations in order to evaluate the hyperfine interaction within the electronic states, correlated to the ground K($4^2S$)+Cs($6^2S$) atomic asymptote. These calculations have shown a significant dependence of the radial matrix elements (both $A^{\mathrm{K}}(R)$ and $A^{\mathrm{Cs}}(R)$) on the internuclear distance $R$ and this made us think that a LIF experiment at higher resolution than in Refs.~\cite{Ferber:2009,Ferber:2013} could provide an evidence for this theoretical prediction. Motivated by these predictions we recorded LIF spectra of the \cstate--\astate transition at the highest achievable spectral resolution, measured the dependence of the HFS splitting on $v_a$ and then compared with the theory. Preliminary analysis showed that the accuracy of the \emph{ab initio} HFS data from Ref.~\cite{Oleynichenko:2020} is very high, but at some points admits further improvements. Since it has been demonstrated that inner-core electronic shell correlations do not affect the shape of hyperfine interaction parameters as functions of $R$, causing a simple shift of these functions, it seems reasonable to undertake new series of calculations, focusing on an extremely accurate description of valence/subvalence correlation effects essential for smoothing the $R$-dependence of hyperfine interactions.

Additional motivation for the present study is the unsufficient coverage of the experimental data on the lowest vibrational levels of the \astate state in Ref.~\cite{Ferber:2009,Ferber:2013}. At that time the LIF was analyzed following excitation of the \Bstate state with diode lasers, available in the laboratory, and the \Estate state was excited using a dye laser with Rhodamine 6G. Under these conditions only few transitions to $v_{a}=0,1, \mathrm{and}~ 2$ were recorded. For the description of the near asymptotic levels in \cite{Ferber:2009,Ferber:2013} this gap in the data set was not crucial, but it definitely should influence the accuracy of potential as a whole, especially regarding the accuracy of the dissociation energy value. In the present experiment we aimed to fill this gap. The short range repulsive part of the \astate state PEC above the dissociation limit was recently determined from bound-free LIF transitions \cite{Krumins:2022}. It was shown that this part of the potential cannot be fixed using trivial extrapolation of the experimental positions of bound and quasi-bound (Feshbach resonances) levels. We will account for the the repulsive part from~\cite{Krumins:2022} while refining the PEC of the \astate state.

The paper is structured as follows. Section \ref{Sec:Exp} contains details on the experimental procedure and description of the recorded LIF spectra, which provide a variety of new data on the \astate state (Section~\ref{Sec:Data}) including the HFS behavior (Section~\ref{Sec:obsHFS}). In Section~\ref{exphfs}  we present a simplified coupled channels (CC) approach used to model the experimental observations. In Section~\ref{abhfs} a brief description of the new \emph{ab initio} calculations is presented. Section~\ref{Sec:fit} describes the refinement of the \astate state potential based on a compilation of all present experimental data and the data from~\cite{Ferber:2009, Ferber:2013}. The last Section~\ref{Sec:hfs} is devoted to semi-empirical analysis of the HFS, based on the experimental observations, \emph{ab initio} data from~\cite{Oleynichenko:2020} and the results of the present calculations presumably providing an improved description of the $R$-dependence of the electronic HFS matrix elements in KCs. As supplementary materials to the paper we provide tables in electronic format with all experimental frequencies (Table I) and newly derived \astate state potential (Table II) and HFS coupling functions (Table III and IV).

\section{Experiment}

\subsection{Experimental set-up}
\label{Sec:Exp}

LIF spectra of the \cstate$\rightarrow$\astate transition excited from the ground \Xstate state were recorded with FT spectrometer IFS125-HR (Bruker) using InGaAs detector. The KCs molecules were produced in a linear heat-pipe, which was heated to about 300$^\circ$ C. For excitation the radiation of a single-frequency Equinox/SolsTis (MSquared) Ti:Sapphire laser was used with a typical line width about 1 MHz and a power of 500 - 600 mW before the entrance window of the heat-pipe. The wave meter WS7-HighFinesse was used for active stabilization of the laser frequency. Numerous LIF spectra were recorded at spectral resolution 0.0063 \cmm, highest achievable by the spectrometer. Exploiting such high resolution, the scanning velocity of the spectrometer's moving mirror was set rather low. At these conditions, in order to achieve sufficient signal-to-noise ratio (SNR), the acquisition time for each spectrum was typically about 8 hours.

In the present experiment we aimed to excite the F$_2$ ($J_{c}=N_{c}$) levels in the \cstate state of $^{39}$K$^{133}$Cs. Overall we have recorded at the highest resolution 17 doublet \emph{P,R} progressions with $N_{a}  = J_{c} \pm 1$ covering \astate state rovibronic levels with $v_{a} \in [0, 27]$ and $N_{a} \in [24, 90]$. Detailed information on the \cstate state was obtained from the recent study \cite{Kruzins:2021}. It was very useful to select the laser frequencies for excitation of a particular rovibronic $v_{c}$, $J_{c}$ levels of the \cstate state, taking into account the strengths of LIF transitions to the \astate state. Selection of different upper state $v_{c}$-levels provided different Franck-Condon factors (FCFs) distribution for the recorded $c-a$ progressions, thus a sufficient amount of transitions with good signal-to-noise ratio (SNR) was ensured for every $v_{a}$. In all spectra along with the $c-a$ progressions in KCs, the K$_2$ $A^1\Sigma^+ \rightarrow X^1\Sigma^+$ transitions were also recorded.

The LIF is collected in a direction opposite to the propagation of the laser beam and this leads to a reduction of the Doppler width of the molecular lines. Indeed, the full-width-half-maximum (FWHM) of the recorded K$_2$ LIF lines to the singlet $X^1\Sigma^+$ state (without HFS) is about 0.011 \cmm, which is less than the expected Doppler width of K$_2$ (0.022 \cmm\ at 10000 \cmm) and is determined mainly by the instrumental apparatus function. We assume that LIF transitions to \Xstate of KCs should be even narrower -- about 0.009 \cmm, which characterizes our actual spectral resolution. The laser frequency was always tuned to the maximum of the fluorescence signal, which should be very close to the center of the molecular line. Small shifts are possible, but they will result only to an overall shift of the particular LIF progression (due to the Doppler effect). Because the frequency range of the fluorescence is narrow (about 200 \cmm) this shift will be almost equal for all frequencies, thus the information on the separation between the \astate state levels will be unaffected.

\begin{figure*}
  \centering
  \epsfig{file=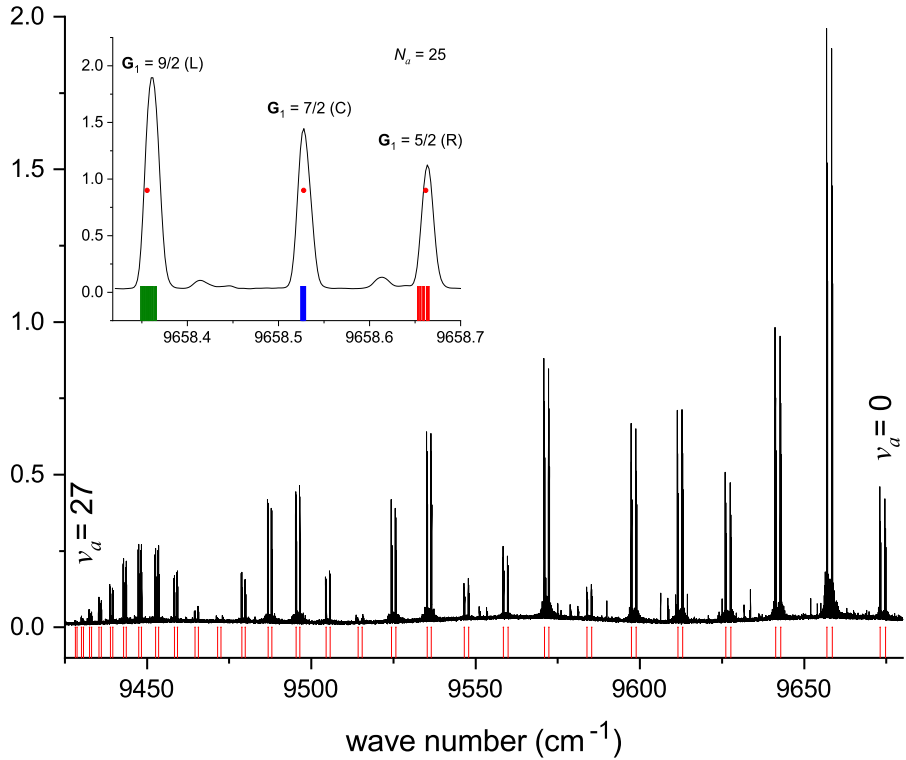,width=0.7\linewidth}
 \caption{$c^3\Sigma^+ \rightarrow a^3\Sigma^+$ doublet $P,R$ progression from $v_{c}$ = 18, $J_{c} = N_{c}=26$ level excited with laser frequency 13373.0989 \cmm. The spectrum is cut off from both sides for better visibility. In the inset a zoom of an $R$-transition (to $N_a$ = 25, $v_a$ = 1) is shown. The red points show the calculated positions of the $G_1$ components and the bars indicate the line positions according to the matrix model from Ref.~\cite{Kasahara:1996} with $A^{\mathrm{Cs}}$ and $A^{\mathrm{K}}$ fixed to the atomic values used in Refs.~\cite{Ferber:2009,Ferber:2013}.}
 \label{spectr}
\end{figure*}

An example of the high-resolution LIF spectrum corresponding to the $c^3\Sigma^+(v_{c},J_{c}=N_{c})\rightarrow a^3\Sigma^+(v_{a},N_{a}=J_{a}= J_{c} \pm 1)$ transition is presented in Fig.~\ref{spectr}. The $P, R$ progression originates from $v_{c}=18$, $J_{c} = 26$ level. Due to the HFS of the \astate state each line is split into three groups, see the inset of Fig.~\ref{spectr}.

\subsection{Experimental $v_{a}$-dependence of the HFS splitting in the \astate state}
\label{Sec:obsHFS}

The observed triplet HFS structure of the \astate state can be understood within the framework of the Hund's coupling $b_{\beta S}$ scheme~\cite{Kasahara:1996}, where the total electron spin ($S =1$) couples with the nuclear spin of Cs ($I_{\mathrm{Cs}} = 7/2$) leading to intermediate angular momentum $\bm{G_1}$ with quantum numbers $G_1= 9/2$ (left group, L in Fig~\ref{spectr}); 7/2 (central group, C), and 5/2 (right group, R). Then $\bm{G_1}$ couples with the nuclear spin of K with $I_{\mathrm{K}} = 3/2$ leading to $\bm{G_2}$, which, finally, couples with $\bm{N}$ to form the total angular momentum $\bm{F}$. The expected splitting between the $G_2$ components of each $G_1$ group is too small when compared with the resolution of the instrument and the residual Doppler width of the lines, therefore most of the analysis in this paper accounts only for the splitting between the $G_1$ components. In fact, when we describe the experimental data by a particular $G_1$ component we mean the group of nearly degenerated $F$ levels around it. The possible importance of the splitting due to K nuclear spin will be discussed at the end of the paper.

\begin{figure}
  \centering
  \epsfig{file=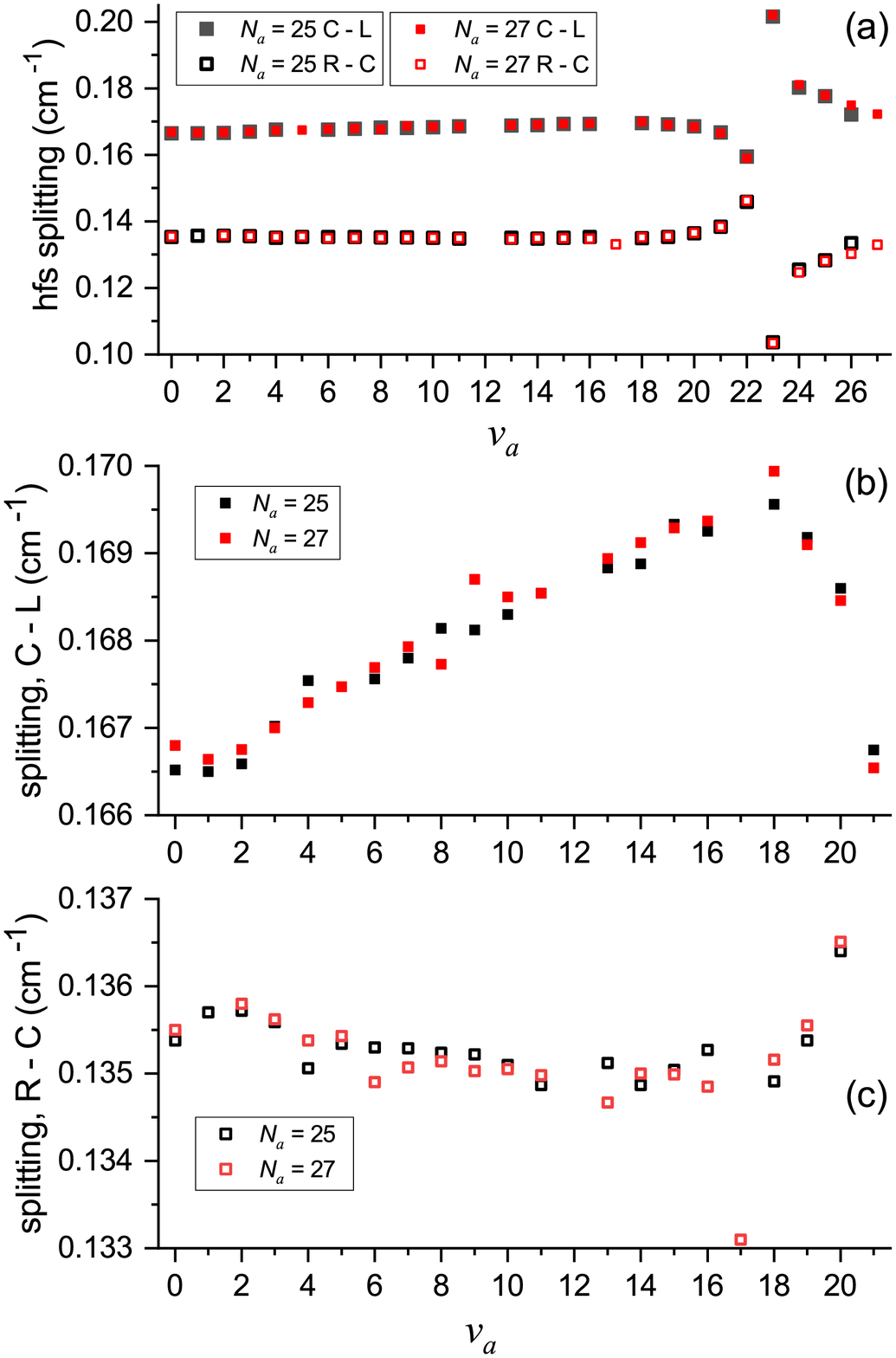,width=\linewidth}
\caption{Example of HFS splitting for $N_{a}=25$ and $27$ as a function of $v_{a}$. In subplot (a) the splitting between the central and left ($G_1=7/2 - G_1=9/2$), as well as between the right and central ($G_1=5/2 - G_1=7/2$) components in a broad scale is shown. The observed avoided crossing for $v_a$ is due to a strong local perturbation. (b) and (c)- respective dependencies in a zoomed energy scale. A weak perturbation can be seen also for $v_a=17$, $N_a=27$ as a sudden drop in the R--C splitting.}
\label{split_zoom}
\end{figure}

As one can see from Fig.~\ref{spectr}, the relative positions of the calculated left and right components are shifted with respect to the experimental one. The applicability of the model from Ref.~\cite{Kasahara:1996} and also the simplified model given in Sec.~\ref{exphfs} may be assessed by the ratio between the splittings (R $-$ C)/(C $-$ L). If interaction with K nuclear spin is neglected the theory says that this ratio should be $3.5/4.5\approx0.778$. For $v_{a}<17$ the interaction with the \Xstate state can also be neglected, so the same ratio should be observed in the experiment. In fact, the experimental ratio for low $v_{a}$ is around $0.81$ and gradually decreases to $0.79$ for $v_{a}=18$. This observations may indicate that the HFS model from Ref.~\cite{Kasahara:1996} is insufficient to explain the HFS within the resolution of the experiment. In Fig~\ref{split_zoom}a one can see the typical splitting between the $G_1$ components of the \astate state as a function of $v_a$. The data come from the LIF progression with $J_{c} = 26$ and the separation between the central and the left ($G_1=7/2 - G_1=9/2$) and the right and the central ($G_1=5/2 - G_1=7/2$) components is shown for both $P$ and $R$ lines. Up to approximately $v_{a}=18$ the splitting between the components depends weakly on the vibrational quantum number $v_a$. At higher $v_{a}$ the interaction with the levels of the \Xstate state increases and one observes initially a gradual change of the position of the central, $G_1=7/2$ component, which moves towards the left ($G_1=9/2$) one. Above $v_{a}=22$ the $a-X$ mixing is very strong (especially for $v_a=23$) and the regular behavior of the splitting is lost. In the scale of Fig~\ref{split_zoom}a one cannot judge about the possible $v$-dependence of the splitting below $v_{a}=18$. This can be followed in Fig~\ref{split_zoom}b and \ref{split_zoom}c where the two splittings ($G_1=7/2 - G_1=9/2$) and ($G_1=5/2 - G_1=7/2$) are shown separately in a zoomed scale. The $v$-dependence of the splitting between the central and the left groups is clearly seen. It is weak (with a span of about 0.003 \cmm) and this explains why it could not be observed in the previous studies \cite{Ferber:2009, Ferber:2013}, where the spectra were recorded at somewhat lower resolution (typically 0.03 \cmm). Seemingly it is the left, $G_1=9/2$ component which position is mostly affected, because the distance between the right and the central components changes to a much smaller extent. However, one can not exclude that the central and the right components are shifted by the same amount and therefore their separation remains almost unchanged.

\begin{figure}
\epsfig{file=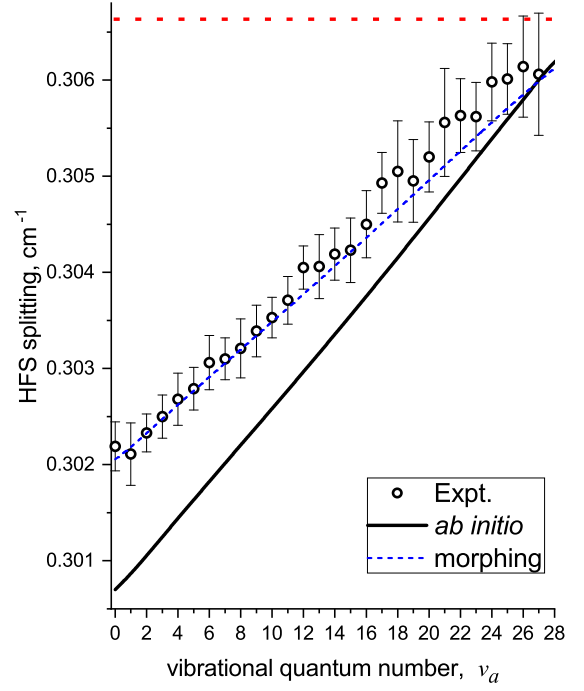, width=\linewidth}
  \caption{The experimental and calculated HFS splitting between the right and the left components ($G_1=5/2 - G_1=9/2$) of the \astate state as a function of the vibrational quantum number $v_{a}$. The experimental data are averaged over all observed rotational quantum numbers $N_{a}$ at high resolution (denoted with black dots in Fig.~\ref{datasets}b). The \emph{ab initio} data were calculated as $\big(F_4(F_4+1)-F_3(F_3+1)\big)\langle v_{a}|A^{\mathrm{Cs}}_{1-1}(R)|v_{a} \rangle$ using the $A^{\mathrm{Cs}}_{1-1}(R)$ function evaluated in Sec.~\ref{abhfs} (see also Fig.~\ref{fig:cs-hfs}), and the Cs atomic quantum numbers $F_3=3$ and $F_4=4$. The dashed blue line corresponds to the morphing function $A^{\mathrm{mor}}_{1-1}(R)=0.77\cdot[A^{\mathrm{Cs}}_{1-1}(R) - A^{\mathrm{Cs}}/2] + A^{\mathrm{Cs}}/2$, where $A^{\mathrm{Cs}}/2$=0.03833 \cmm\ is half of the HFS constant of the Cs atom~\cite{Cshfsexp}. The horizontal red dotted line denotes the Cs atom splitting $F_4-F_3$=0.306633 \cmm.}
  \label{split_ave}
\end{figure}

Such routine analysis of the HFS splitting has been carried out for all progressions recorded at the ultimate resolution of the FT spectrometer and similar behavior of the HFS has been observed. According to our analysis the splitting between the $G_1=9/2$ and $G_1=5/2$ components (R--L) remains virtually independent of the rotational quantum number $N_{a}$, hence it was averaged over all $N_{a} \in [24,90]$. The average change of the R--L splitting is shown in Figure~\ref{split_ave}. The presented monotonic increase proofs that the R--L splitting is not affected by $a\sim X$ mixing, even for $v_a = 23$.

Apparently the conventional assumption that the molecular HFS parameters may be treated as $R$-independent constants,  which are close to their atomic counterparts, fails at such resolution and the main goal of this paper is to check whether the recent \emph{ab initio} calculations (Ref.~\cite{Oleynichenko:2020} and also this study) may offer a solution to the problem.

\subsection{Upgraded line list for the \astate state}
\label{Sec:Data}

\begin{figure}
  \centering
  \epsfig{file=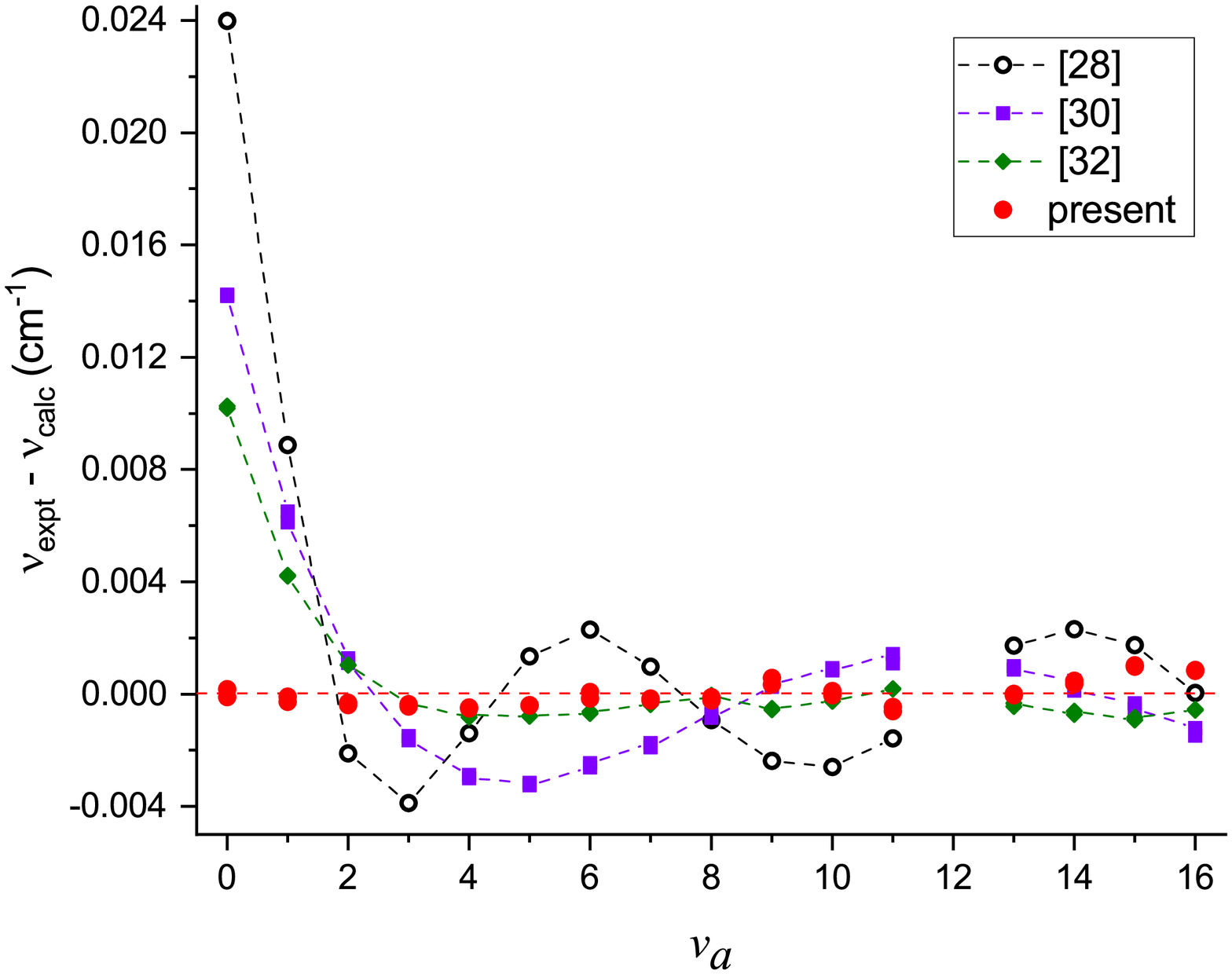,width=\linewidth}
  \caption{Deviations between the experimental and calculated frequencies for a progression from $v_{c}$=18, $J_{c}$=26 level to $v_{a}$-levels with $N_a$ = 25, 27 of \astate potentials from Ref.~\cite{Ferber:2013} (black open circles), Ref.~\cite{Schwarzer:2021} (violet solid squares), Ref.~\cite{Krumins:2022} (green diamonds), and this study (red solid dots).
  The available PECs~\cite{Ferber:2013, Schwarzer:2021, Krumins:2022} were all constructed using the previous experimental data given in the Supplementary Materials of Refs.~\cite{Ferber:2009, Ferber:2013} (see also Fig.\ref{datasets}a)} \label{comp-pots}
\end{figure}

As already mentioned in the Introduction, the data set for the \astate state from \cite{Ferber:2009, Ferber:2013} poorly represents the first vibrational levels (very few observations for $v_{a}=0-2$). With the Ti:Sapphire laser it was possible to excite LIF from a different electronic state, the \cstate one, where the FCFs for observation of these levels turned out to be more favorable. We checked the extrapolation properties of the PEC from \cite{Ferber:2013} and we figured out that it needs some refinement. In Fig.~\ref{comp-pots} one can see the differences between the experimental line frequencies of the progression, originating from $v_{c}=18$ and $J_{c}=26$, and the predicted ones based on the previous semi-empirical PECs~\cite{Ferber:2013, Schwarzer:2021, Krumins:2022} using an adiabatic (a single channel) model. The comparison is for data below $v_{a}=17$ since the levels above are affected by the HF coupling to the \Xstate state, see Fig.~\ref{split_zoom}. The deviations for $v_{a}<2$ exceed the experimental accuracy and within the current study we are going to propose an improved version of the \astate state potential.

\begin{figure*}
  \centering
   \epsfig{file=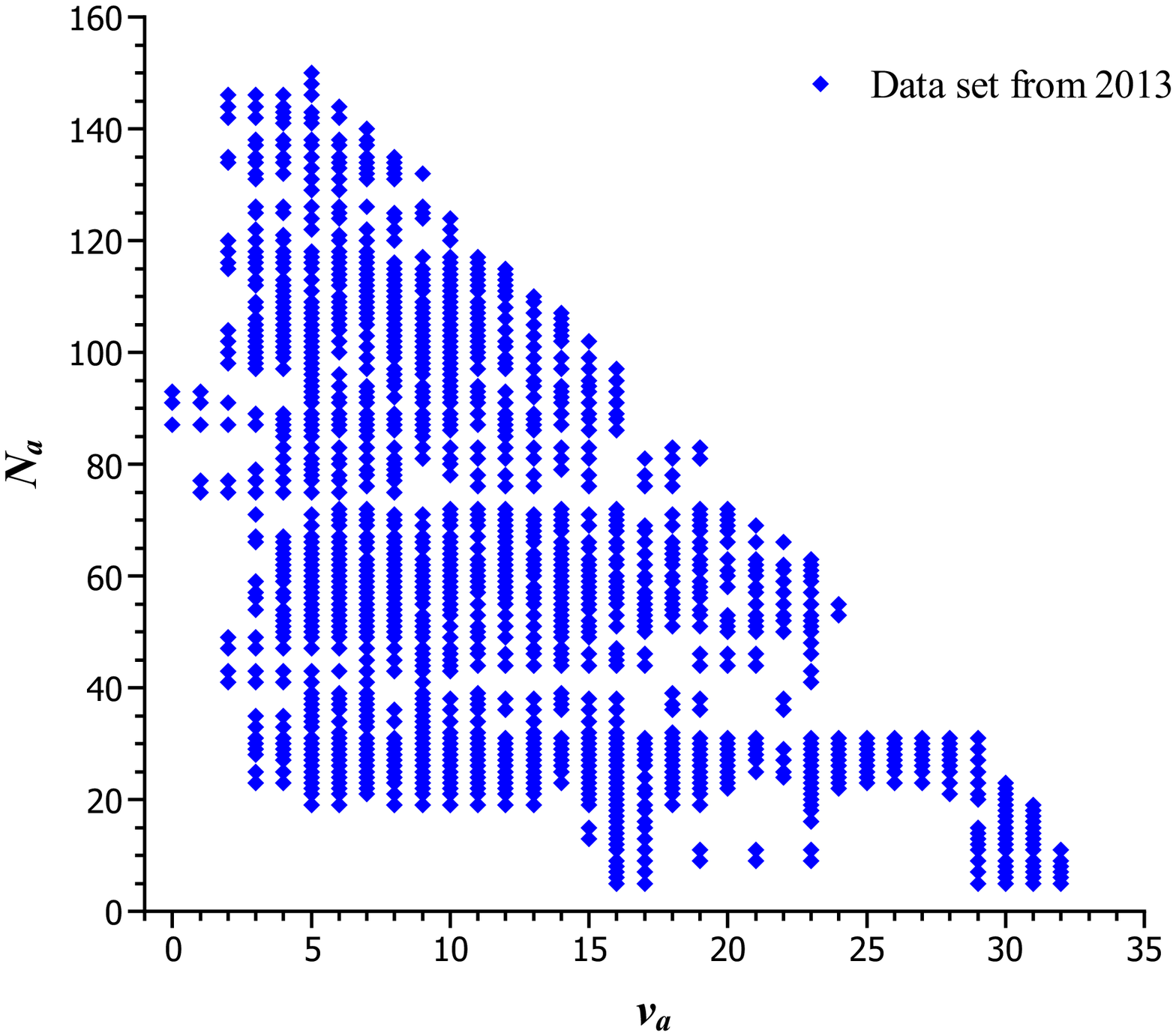,width=0.47\linewidth}a
   \epsfig{file=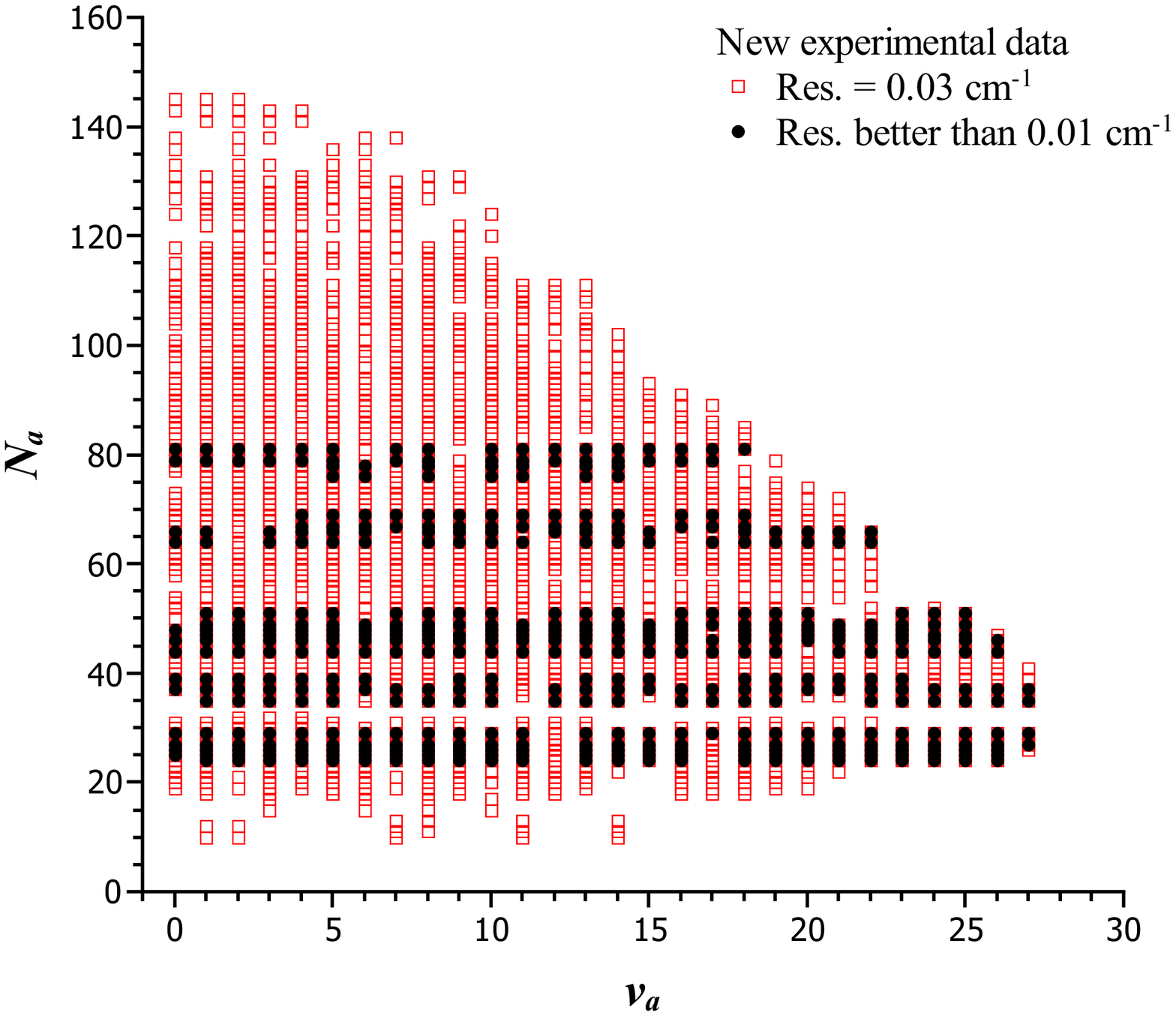,width=0.47\linewidth}b
  \caption{Data sets: (a) 2103 lines from 2013 \cite{Ferber:2013} (blue diamonds) and (b) 4500 lines from Ref.~\cite{Kruzins:2021} and this study (red open squares). With black solid circles the 570 high resolution (better than 0.01 \cmm) data points are denoted.} \label{datasets}
\end{figure*}

In Fig.~\ref{datasets} we compare the data sets used to build the PEC from 2013 \cite{Ferber:2013} (a total of 2103 lines, Fig.~\ref{datasets}a) with the new lines added in this study plus the data from Ref.~\cite{Kruzins:2021} (4500 lines, Fig.~\ref{datasets}b). One can clearly see that the new measurements filled the gap for the lowest vibrational levels. With black solid dots the 570 high resolution lines (better than 0.01 \cmm) are denoted.

\section{Modelling}
\label{Sec:Model}

\subsection{The simplified treatment of the \astate hyperfine structure}
\label{exphfs}

It is assumed that the HFS of the \astate state is only due to interaction between the electronic spin $\bm{S}$ and the nuclear spin of Caesium $\bm{I}_{\mathrm{Cs}}$ and that this interaction can be separated from the nuclear rotation. The first assumption is justified because the splitting due to K nuclear spin is hardly visible, at most as a broadening of the triplet lines at the highest resolution of the instrument. The second assumption is also reasonable since virtually no change of the HFS is observed for lines with different rotational quantum numbers. Thus, we assume that HFS is induced only by the interaction of the magnetic moment of Cs nuclei with the electronic subsystem:

\begin{eqnarray}
H^{\rm{mg}}=A^{\mathrm{Cs}}\bm{I}_{\rm Cs}\bm{S}
\end{eqnarray}

Within the Hund's case $b_{\beta S}$ the HFS structure of the $F_2$ component of the \astate\ may be described by a simple four coupled-channel model (K nuclear spin excluded). The total HFS matrix for this model is shown in Figure~\ref{Hamiltonian_Table}. Without HFS interactions, the energy structure is governed by the adiabatic PECs $U_X(R)$ and $U_a(R)$. The \astate state levels split into three $G_1$ HFS components and the splitting is given by the diagonal matrix element of $A^{\mathrm{Cs}}_{1-1}$ for the $a^3\Sigma^+_1$ state:

\begin{equation}
A^{\mathrm{Cs}}_{1-1}[G_1(G_1+1)-I(I+1)-S(S+1)]/2 \mbox{ ,}
\label{eq1}
\end{equation}

\noindent where $S=1$, $I=7/2$ while $G_1=5/2$, $7/2$ and $9/2$. The interaction between the $G_1=7/2$ component with the singlet \Xstate state levels is described by the off-diagonal matrix element:

\begin{equation}
A^{\mathrm{Cs}}_{0^+-1}\sqrt{G_1(G_1+1)} \mbox{ .}
\label{eq2}
\end{equation}

As one can see from Figure~\ref{Hamiltonian_Table}, the interactions couple only two of the channels (\Xstate and $a^3\Sigma^+_{G_1=7/2}$). The HFS matrix for the other $G_1$ components of the \astate state is diagonal so one can approximate the singlet-triplet interaction by the $2\times 2$ coupled-channels model. Let us remind that in the list of experimental data the frequency of the transition namely to the $G_1=7/2$ component is given.

\begin{figure*}[!ht]
\renewcommand\arraystretch{1.5}
\caption{The simplified HFS matrix for the present coupled-channel problem. The diagonal kinetic energy terms $-\hbar^2/(2\mu)(d^2/dR^2)$ and the rotational energy terms $\hbar^2J(J+1)/(2\mu R^2)$ are omitted for sake of clarity.}
\begin{equation*}
\begin{blockarray}{ccccc}
 &{^1\Sigma^+} &  {^3\Sigma^+_{G_1=7/2}} &  {^3\Sigma^+_{G_1=9/2}} & {^3\Sigma^+_{G_1=5/2}} \\
\begin{block}{c(cccc)}
 {^1\Sigma^+} \;\;           &U_X(R)\;\; &\frac{3\sqrt{7}}{2}A^{\mathrm{Cs}}_{0^+-1}(R)      &                    &                   \\
 {^3\Sigma^+_{G_1=7/2}} \;\; &\frac{3\sqrt{7}}{2}A_{0^+-1}^{\mathrm{Cs}}(R)\;\;&U_a(R)-A^{\mathrm{Cs}}_{1-1}(R)&                    &                    \\
 {^3\Sigma^+_{G_1=9/2}} \;\; &           &                 &U_a(R)+7/2A^{\mathrm{Cs}}_{1-1}(R)&                    \\
 {^3\Sigma^+_{G_1=5/2}}\;\;  &           &                 &                    &U_a(R)-9/2A^{\mathrm{Cs}}_{1-1}(R)\\
\end{block}
\end{blockarray}
\end{equation*}
\label{Hamiltonian_Table}
\end{figure*}

In this study the CC system of radial Schr\"odinger equations modeled by the Hamiltonian matrix from Fig.~\ref{Hamiltonian_Table} is solved by the Fourier-Grid method \cite{FGH} by a routine described in more details in Ref.~\cite{Rb2depert}. In the present case the Hamiltonian matrix has been calculated in a non-equidistant grid of 300 points in the interval $R\in[2.9, 40]$ \AA\ by using a mapping procedure.

\subsection{\emph{Ab initio} calculation of HFS parameters}
\label{abhfs}

To construct the CC model one has to provide the electronic parameters $A^{\mathrm{Cs}}_{1-1}$ and $A^{\mathrm{Cs}}_{0^+-1}$ as function of the internuclear separation $R$. The required $R$-dependent matrix elements have been already obtained in Ref.~\cite{Oleynichenko:2020} within the four-component relativistic multi-reference Fock space coupled cluster methodology. A rigorous treatment of core and core-valence relaxation and correlation contributions to $A$ parameters within the employed approach had demonstrated the practical independence of these contributions of $R$~\cite{Oleynichenko:2020}. For this reason it seems more practical to employ the relativistic pseudopotential approximation, which completely ignores the mentioned core contributions, and  then simply to shift the resulting hyperfine radial functions to fit the asymptotic (atomic) values of HFS constants. Electronic wavefunctions in the vicinity of the atomic nucleus, which determines the magnetic dipole hyperfine interaction and which is not reproduced  directly within the pseudopotential approach, were obtained within the a posteriori restoration technique (see~\cite{Titov:2005} and references therein). This approach had been already used to study hyperfine interaction in the LiRb and LiCs molecules~\cite{Bormotova:2020}.

The interaction of the nuclear magnetic dipole moment $\bm{\mu} = \mu \mu_N \bm{I}/I$ (where $\mu$ stands for the value of the nuclear magnetic moment in nuclear magnetons, $\mu_N$, and $\bm{I}$ is the nuclear spin) with electrons can be represented by the expression (see e.g. \cite{Norman:2018}):

\begin{equation}\label{eq:hfs-el-nuc}
H^{\rm mg} = \bm{\mu} \cdot \sum_i \frac{ [ \bm{r}_i \times \bm{\alpha}_i ] }{r_i^3}
\end{equation}

\noindent where $\bm{\alpha}_i$ are the Dirac alpha matrices and $\bm{r}_i$ is a radius vector of the $i$-th electron with respect to the center of the chosen nucleus. The spin-spin interaction between nuclei is typically three order of magnitude less than $H^{\rm mg}$, and thus the HFS induced by the K and Cs nuclei can be treated separately.

In the present work the electronic structure of KCs was studied using the relativistic Fock space coupled cluster method with single and double excitations (FS-RCCSD)~\cite{Kruzins:2021}. 10 core electrons of K and 46 electrons of Cs were simulated using the semi-local shape-consistent relativistic pseudopotentials~\cite{Mosyagin:2010, RPP-website}. The basis sets used to describe valence/subvalence shells of K and Cs comprised $[7s7p6d4f2g]$ and $[7s7p6d4f3g1h]$ contracted Gaussian functions ~\cite{Zaitsevskii:2017, Kruzins:2021}, respectively. Model space Slater determinants were obtained by distribution of two valence electrons over the 122 lowest-energy molecular spinors of the closed-shell state of KCs$^{2+}$ ion considered as the Fermi vacuum. This model space is more than order of magnitude larger than those employed in~\cite{Oleynichenko:2020}; this guarantees the smoothness of the obtained hyperfine radial functions. Numerical instabilities due to the presence of intruder states inevitable for such vast model spaces were suppressed by using adjustable energy denominator shifts \cite{Zaitsevskii:2017}. The real simulation of imaginary shifts was employed (formula (8) in \cite{Oleynichenko:2020}). The shift parameter $S_K = -0.6$ a.u. was used for all excitations in the target two-valence-particle Fock space sector. Hyperfine interaction matrix elements calculated using the finite-field technique \cite{Zaitsevskii:2018, Oleynichenko:2020} were found to be very stable with respect to the variation of the denominator shift parameters. All coupled cluster and finite-field calculations were performed using the EXP-T software \cite{Oleynichenko-EXPT, EXPT-website}. The DIRAC19 program package~\cite{DIRAC19, Saue:2020} was used to solve relativistic Hartree-Fock equations and to transform the molecular integrals. Restoration of the valence wavefunction in the core region and evaluation of magnetic dipole hyperfine interaction integrals were performed using the OneProp program~\cite{Skripnikov:2011, Skripnikov:2015}.

\begin{figure}
\epsfig{file=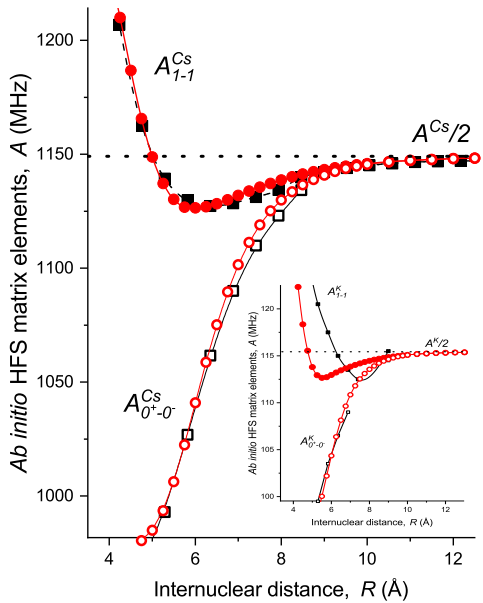,width=0.8\linewidth}
\caption{The \emph{ab initio} $1-1$ and $0^+-0^-$ electronic matrix elements $A^{\rm Cs}$, $A^{\rm K}$ of magnetic dipole hyperfine interaction for the \Xstate and \astate states of $^{39}$K$^{133}$Cs induced by the $^{133}$Cs and $^{39}$K nucleus, respectively. Open and solid black symbols denote the results of previous four-component FS-RCCSD calculations~\cite{Oleynichenko:2020}. The horizontal dots lines show the corresponding values of the experimental atomic HFS constants.}
\label{fig:cs-hfs}
\end{figure}

The resulting $A^{\rm Cs}(R)$ and $A^{\rm K}(R)$ radial functions ($1-1$ and $0^+-0^-$) for the $^{39}$K$^{133}$Cs isotopologue are depicted in Fig.~\ref{fig:cs-hfs}. The values of nuclear magnetic moments, $\mu(^{133}Cs) = 2.582025\ \mu_N$ ($I=7/2$) and $\mu(^{39}K) = 0.39147\ \mu_N$ ($I=3/2$), were taken from \cite{Stone:2005}. The results of previous four-component FS-RCCSD calculations \cite{Oleynichenko:2020} are also given on Fig.~\ref{fig:cs-hfs}. Note that all curves were uniformly shifted to fit the corresponding atomic values ($A^{\rm Cs}/2 = 1148.1$ and $A^{\rm K}/2 = 115.4$ MHz) at the large distance. The reliability of the present \emph{ab initio} HFS  functions in the ``valence'' region ($R > 5$\ \AA) seems to be higher than in the previous research due to the use of more flexible basis sets including high-$l$ basis functions and much larger model space.

\subsection{Direct-potential-fit of the \astate state}
\label{Sec:fit}

In the previous works~\cite{Ferber:2009, Ferber:2013} the pair of potential curves for the \astate and \Xstate states were simultaneously fitted to the experimental data available at that time (see the supplementary materials to Ref.~\cite{Ferber:2013}) assuming the constant hyperfine radial functions $A^{\mathrm{K}}(R)$ and $A^{\mathrm{Cs}}(R)$ (fixed to their asymptotic atomic values). Since the new experimental data require refinement of the \astate state potential from Ref.~\cite{Ferber:2013} only around its minimum, it is possible to simplify the highly sophisticated coupled-channel (CC) fitting procedure applied in Ref.~\cite{Ferber:2009,Ferber:2013}. We chose the point wise form of the PEC because it is flexible and makes it possible to change only a small part of the curve while leaving the rest unaffected. At long internuclear distances the potential was described by the traditional $C_n$ dispersion coefficients, exactly as in Ref.~\cite{Ferber:2013}. The fitting routine is based on a single-channel realization of the Inverted perturbation approach from Ref.~\cite{ipaasen}.

As a first step we combined all the experimental data on the \astate state from Ref.~\cite{Ferber:2009,Ferber:2013} and the present study. In a total we get a list of about 6600 frequencies (see Table I with the list of all frequencies in the Supplementary materials). With the new experimental data we refined the lower part of the PEC by using data only up to $v_{a}=17$, because below this vibrational quantum number the coupling with the ground singlet state may be ignored. Then, we needed to ensure that the new potential leaves the positions of the levels with $v_{a}>17$ unchanged. We calculated the level energies of the \astate state with the potential from Ref.~\cite{Ferber:2013} for $v_{a}\in[18,32]$ for a wide range of $N_{a}\in[1,40]$ (a total of 550 ``deperturbed'' values) and added them as synthetic data with an uncertainty of 0.001 \cmm\ to the experimental data set (with $v_{a}\le17$). We refitted the inner part of the potential again while keeping the long range part exactly as defined in Ref.~\cite{Ferber:2013}. In this way ensured that the new potential will be corrected around its minimum and at the same time it will remain virtually the same for higher $v_{a}$ as in Ref.~\cite{Ferber:2013}. The term energies of the upper states for all experimental frequencies were treated as free parameters -- separate parameter for every LIF progression. In this way we avoided problems with possible excitation of the same upper state levels with slightly off-resonance laser. The synthetic energies of the \astate state were ``converted'' into frequencies by assuming a common upper term with energy equal to zero -- the value of the atomic asymptote.

After few iterations it was possible to reproduce all the experimental frequencies ($v_{a}\leq 17$) plus the synthetic ones (a total of 5928) with a standard deviation of 0.0022 \cmm\ and a dimensionless standard deviation 0.66. As already mentioned, the frequency of the central $G_1=7/2$ component of the experimental HFS triplets is listed in the experimental data. At the same time the synthetic data reflect the deperturbed positions of the triplet state levels. By fitting these two data sets together, the shift of the experimental data from the unperturbed position is included in the fitted term energies of the upper state for the corresponding progressions. This is possible, because at this stage only experimental frequencies to levels with $v_{a}\leq 17$ were fitted, therefore the shift is the same for all lines (no interactions with the \Xstate state). By keeping the long range part of the fitted potential fixed we ensure the proper position of the triplet potential curve with respect to the singlet one.

We believe that the new \astate state potential (Table II of the supplementary materials \cite{EPAPS} ) should have similar quality for higher vibrational levels as the previous one \cite{Ferber:2013} and even superior for lower $v_{a}\leq 17$, see Fig~\ref{comp-pots} and Table I from \cite{EPAPS}. In order to check this we applied the simple CC model from this section (see Figure~\ref{Hamiltonian_Table}), which also served as a tool to check the quality of the $R$-dependence of the hyperfine coupling functions from Ref.~\cite{Oleynichenko:2020} and this study. As already mentioned the new potential at short range is supplemented with the repulsive branch, obtained in \cite{Krumins:2022}, so it is able to reproduce also the observed bound-free transitions.

\begin{figure}
  \centering
  \epsfig{file=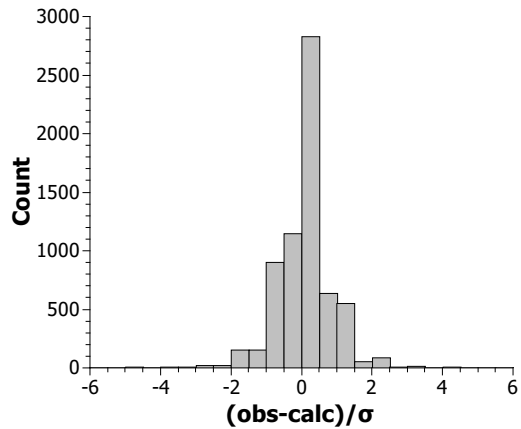,width=0.8\linewidth}
  \caption{Histogram of the residuals of coupled channels fit, normalized by the experimental uncertainty $\sigma$.}
  \label{obs-calc}
\end{figure}

When testing the quality of the new \astate potential the whole experimental data set was used -- a total of 6600 transition frequencies (Table I of  \cite{EPAPS}). The ground \Xstate state potential curve is taken from Ref.~\cite{Ferber:2013}. The program finds the eigenvalues of the two-channels Hamiltonian and then fits the energies of the upper terms (separate value for each progression) in order to match the experimental frequencies. The calculations show an agreement with the experiment with a standard deviation of 0.0026 \cmm\ and a dimensionless standard deviation of 0.81, which is comparable with the quality obtained in the previous study Ref.~\cite{Ferber:2013}. The histogram of the residuals of coupled channels fit, normalized by the experimental uncertainty is presented in Fig.~\ref{obs-calc}. This confirms the usefulness of the two-channels approach and justifies the use of the simple model for the singlet-triplet interaction.

\section{Semi-empirical analysis of the observed hyperfine splitting}
\label{Sec:hfs}

In order to check the quality of the theoretical calculations on the HFS and the performance of the refined potential curve we used the coupled channels model defined in Sec. \ref{abhfs} (Fig.~\ref{Hamiltonian_Table}). The coupling between the \Xstate state and the $G_1=7/2$ component of the \astate state is modeled with the \emph{ab initio} $A^{\mathrm{Cs}}_{0^+-1}(R)$ function from Ref.~\cite{Oleynichenko:2020}, whereas for the splitting between the $G_1$ components we compare the results with the asymptotic (atomic) value of $A^{\mathrm{Cs}}_{1-1}\equiv A^{\mathrm{Cs}}/2$=0.0383~\cmm\ used in Ref.~\cite{Ferber:2009, Ferber:2013} and with the present \emph{ab initio} $A^{\mathrm{Cs}}_{1-1}(R)$ molecular coupling function.

\begin{figure*}
  \centering
   \epsfig{file=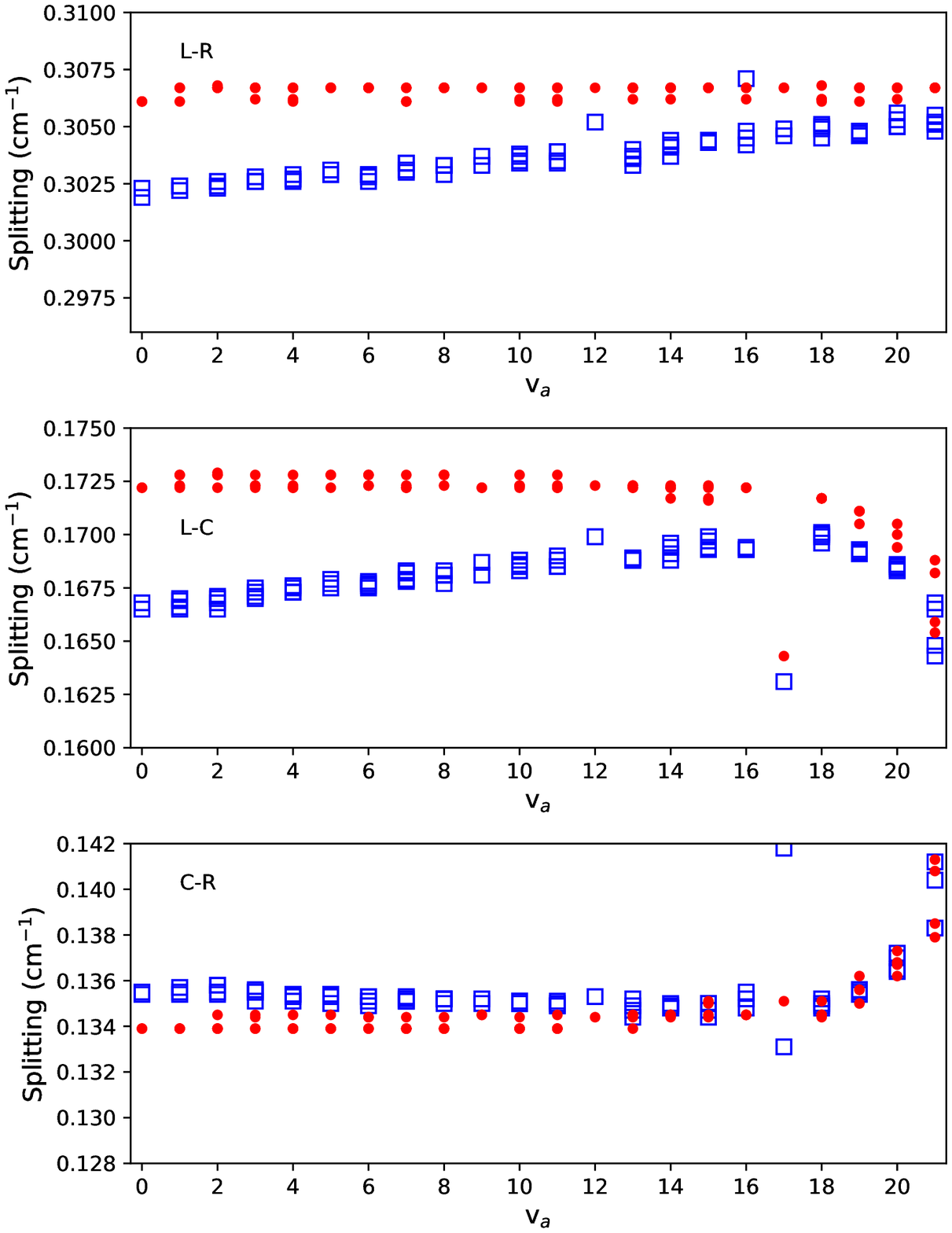,width=0.47\linewidth}a
   \epsfig{file=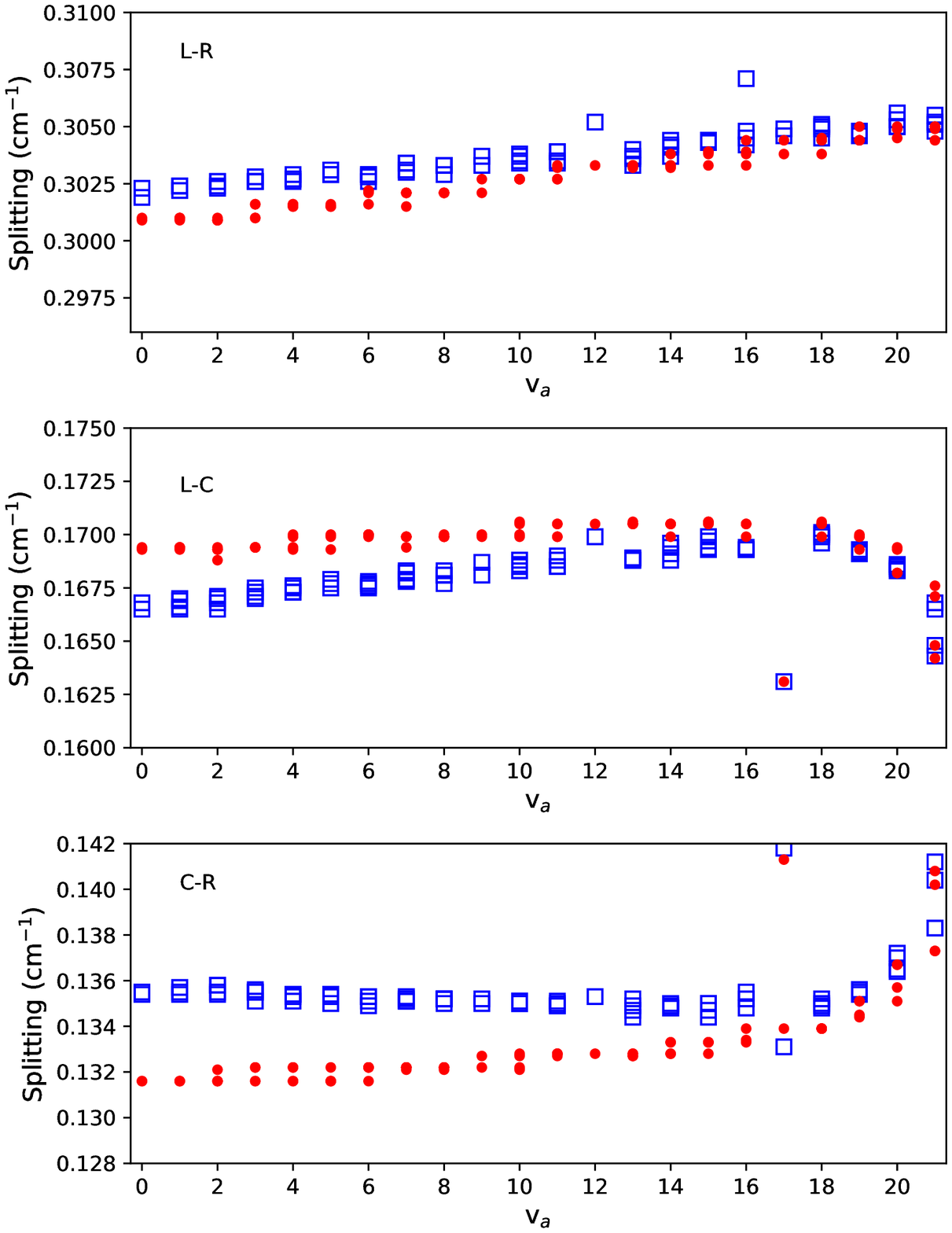,width=0.47\linewidth}b
  \caption{Comparison between the experimental (blue open squares) and the calculated by the coupled channels FGH method (red dots) splittings between the three hyperfine components of the \astate state levels with $N_{a}=$ 25, 27, 46 and 48. In part (a) the CC calculations are carried out with constant coupling function $A^{\mathrm{Cs}}_{1-1}(R)=0.0383$ \cmm, in part (b) the function is taken from Ref.~\cite{Oleynichenko:2020}.}
  \label{AR-comp}
\end{figure*}

In Fig.~\ref{AR-comp} we show the results obtained for the progression from $J_{c}=26$ (similar to those discussed already in connection with Fig.~\ref{split_zoom}) and $J_{c}=47$ . According to the model adopted in this study the splitting between the $G_1=9/2$ and $G_1=5/2$ is only due to the diagonal correction to these \astate state components. Therefore this splitting may be used to directly test the $A^{\mathrm{Cs}}_{1-1}(R)$ coupling function from Ref.~\cite{Oleynichenko:2020} and this study. The position of the central component $G_1=7/2$ is affected also by the \Xstate state levels, and thus by the $A^{\mathrm{Cs}}_{0^+-1}(R)$ function, and also by the quality of the PECs of both states.

The separation between the $G_1=9/2$ and $G_1=5/2$ components (R--L, compare with Fig.~\ref{split_ave}) with the constant interaction function is shown in the upper part of Fig.~\ref{AR-comp}a. Only data for $v_{a}<22$ are included. The calculated separation remains constant (within the accuracy of the calculations), which is very different from the observed $v_{a}$-dependence. The situation changes when the newly calculated  $A^{\mathrm{Cs}}_{1-1}(R)$ coupling function is used, see the upper part of Fig.~\ref{AR-comp}b. Although the calculated separation is a little bit too small by about 0.001 \cmm, the trend of narrowing the R--L splitting with diminishing $v_{a}$ is very similar to the experimental observations. The present result is remarkable because it is based on pure \emph{ab initio} function without any fitting to the experimental data. With the $A^{\mathrm{Cs}}_{1-1}(R)$ function from Ref.~\cite{Oleynichenko:2020} very similar results can been obtained, which confirms the reliability of both theoretical approaches.

The separation between the C--L and R--C components is not reproduced properly by the calculations. The $G_1=9/2$ and $G_1=5/2$ components are shifted to smaller frequencies by nearly the same amount, resulting in a proper R-L splitting, but R--C is underestimated and C--L -- overestimated. The result is consistent with the inset in Fig.~\ref{spectr}. This disagreement might be explained also by the improper position of the central component $G_1=7/2$ with respect to the other two. However, at small $v_{a}$ the interaction with the \Xstate is too weak to be responsible for such shifts. By cancelling the coupling between the singlet and the triplet states, only the positions of $v_{a}>17$ are significantly affected. Apparently a more careful analysis of the HFS model is necessary in order to explain the splitting between the observed HFS components of the \astate state.

The simple model used so far tries to explain the observed splitting between the HFS components only by the FC interaction and the positions of the $G_1$ components. Here (as well as in Refs.~\cite{Ferber:2009,Ferber:2013}), the lineshapes and the HFS of the excited states are completely neglected. It is considered that all hyperfine components of the upper electronic state are populated by the exciting laser and therefore all possible hyperfine transitions are observed in fluorescence. Indeed, these assumptions seemed reasonable at lower resolution (where other factors are dominant for the line shape) and they were confirmed by similar studies of other molecules like NaRb, NaCs, LiCs and others \cite{Pashov:2005,Docenko:2006,Staanum:2007} where the HFS appears as independent of the vibrational and rotational quantum numbers of the \astate state and also of` the quantum numbers of the excited level. However, in one case, that of KRb \cite{Pashov:2007}, it was noticed that the HFS changes when the laser is tuned across the Doppler profile of the excitation transition. Of course it may be the case also for the other molecules, but to somewhat less extent.

It is possible to try to model the splitting within the Hund's coupling case b$_{\beta S}$ by a matrix approach taking into account both nuclear spins as explained in detail in Ref.~\cite{Kasahara:1996}. From the same paper we took the expression to calculate also the line intensities for the \cstate $\rightarrow$ \astate hyperfine transitions. It is also possible to add the dipole-dipole interaction between the electron spin and the nuclear spins as in Ref.~\cite{Kato:1993}. In this paper again the spins are assumed to be decoupled from the molecular rotation and the only effect of the nuclear motion comes from the $v$-dependent mean values of the coupling constants A$^{\mathrm{K}}_{v}$, A$^{\mathrm{Cs}}_{v}$ (for the FC interaction) and D$^{\mathrm{K}}_{v}$, D$^{\mathrm{Cs}}_{v}$ (for the dipole-dipole interaction). Here we adopted a linear dependence of this constants on $v$. To perform a fit, we first simulated the line profile of the HFS triplet for given vibrational level $v_{a}$, find the peaks of the three components and compare with the experimental separations for $v_{a}<22$.

\begin{figure*}
  \centering
   \epsfig{file=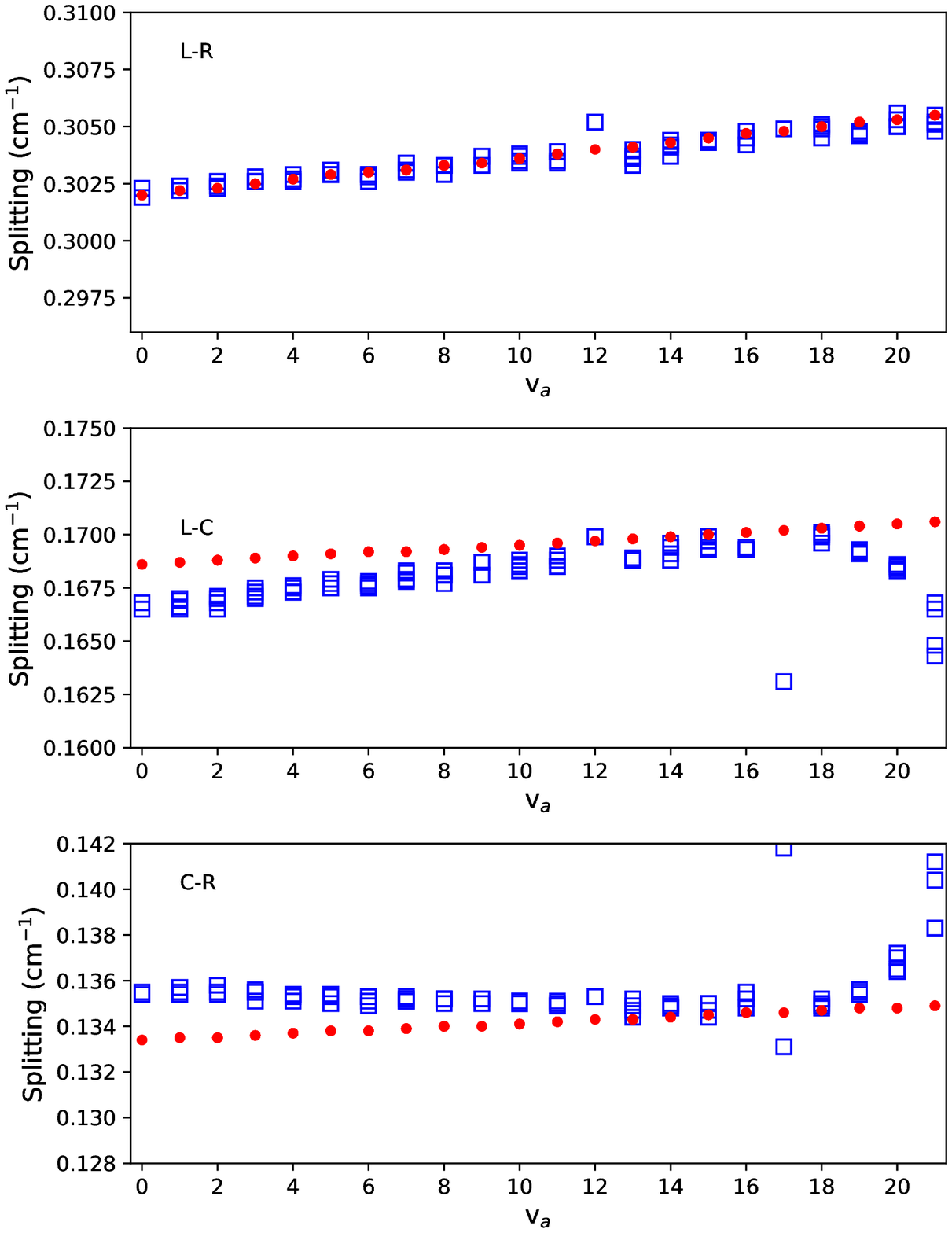,width=0.47\linewidth}a
   \epsfig{file=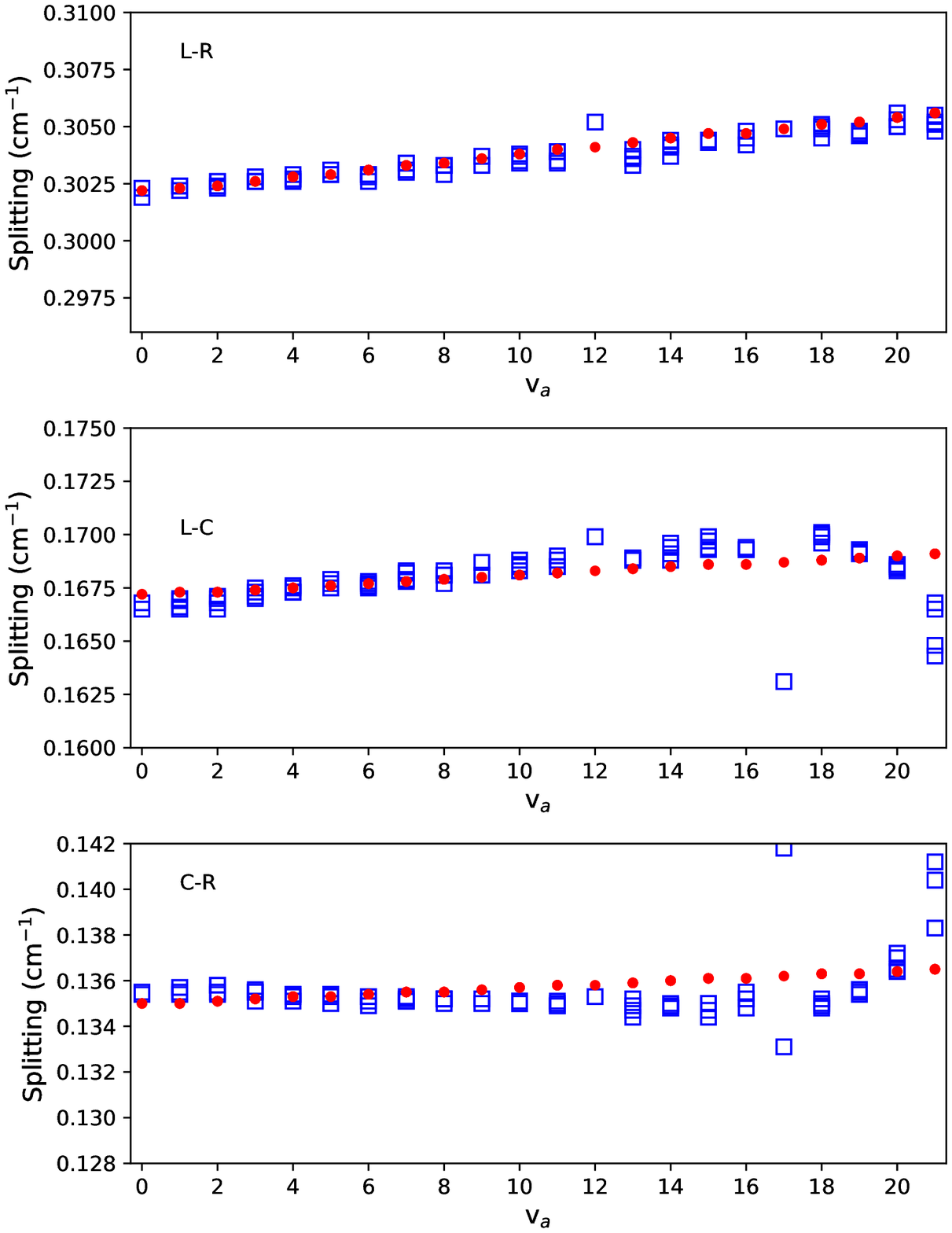,width=0.47\linewidth}b
  \caption{Comparison between the experimental (blue open squares) and the calculated by the matrix method (red dots) splittings between the three hyperfine components of the \astate state levels with $N_{a}=$ 25, 27, 46 and 48. In part (a) the calculations are carried out with A$^{\mathrm{Cs}}_{v}=0.07483(11)+4.2(12)\times10^{-5}v_{a}$ \cmm; in part (b) along with the FC constant A$^{\mathrm{Cs}}_{v}=0.07689(17)+4.1(12)\times10^{-5}v_{a}$ \cmm, the dipole-dipole interaction is added and D$^{\mathrm{Cs}}_{v}=-1.92(24)\times10^{-3}$ \cmm.} \label{A-comp}
\end{figure*}

We performed various fits and figured out that within the present data accuracy the parameters concerning K atom may be kept fixed to their asymptotic value: $A^{\mathrm{K}}_{v}=0.0077$ \cmm\ and $D^{\mathrm{K}}_{v}$ is neglected. By using a linear dependence for A$^{\mathrm{Cs}}_{v}$ and setting D$^{\mathrm{Cs}}_{v}$ to zero we obtained similar representation of the HFS splitting as with the CC model -- the results are shown in Fig~\ref{A-comp}a. The fitted values are A$^{\mathrm{Cs}}_{v}=0.07483(11)+4.2(12)\times10^{-5}v_{a}$ \cmm. Of course, after $v_{a}=18$ we see deviation in the R--C and C--L separations, because in the present model the interaction of the central component with the \Xstate state is neglected. The systematic underestimate of the R--C splitting and the overestimate of the C--L splitting again shows that the FC interaction probably is not the only interaction responsible for the HFS of the \astate state levels. However the agreement between the experiment and the calculation seems to be somewhat better than in the current CC model (compare with the R--C and C--L splittings in Fig.~\ref{AR-comp}b). We found that the reason for this, at least to some extent, is that in the CC model the line profile and the interaction with the K nuclear spin are completely neglected. For example the ratio (C $-$ L)/(R $-$ C) changes from 0.778 for the CC calculation to 0.792 when $A^{\mathrm{K}}_{v}=0.0077$  and a line width of 0.011 \cmm\ are used.

A possible further extension of the present HFS model is to include the dipole-dipole interaction \cite{Kato:1993}. It turns out that it changes the ratio between the R--C and C--L splittings and therefore it is plausible to test whether a better fit may be obtained. The result is shown in Fig.~\ref{A-comp}b. The linear dependence of the FC interaction is A$^{\mathrm{Cs}}_{v}=0.07689(17)+4.1(12)\times10^{-5}v_{a}$ \cmm\ and D$^{\mathrm{Cs}}_{v}=-1.92(24)\times10^{-3}$ \cmm. Including a linear term for the dipole-dipole interaction does not improve the agreement with the experiment. The list of experimental frequencies used to illustrate the HFS splitting in Figs.~\ref{split_zoom},\ref{AR-comp} and \ref{A-comp} is presented in Table V of the Supplementary material \cite{EPAPS}.

\section{Concluding remarks}
\label{Sec:concl}

Within the present study the potential energy curve of the \astate state state in KCs was refined. Compared to the PECs from~\cite{Ferber:2013},~\cite{Schwarzer:2021}, and~\cite{Krumins:2022} the present interatomic potential has better accuracy around the potential minimum. The present dissociation energy (potential depth) $D_{\mathrm{e}}$ = 267.21(1)\cmm is in excellent agreement with $D_{\mathrm{e}}$ value in~\cite{Krumins:2022} in distinction from the values 267.173(10) \cmm\ \cite{Ferber:2013}, and 267.251 \cmm\ in \cite{Schwarzer:2021}. The value of the equilibrium distance $R_{\mathrm{e}}=6.050(2)$ \AA\ remains the same. The present PEC shares the repulsive branch above the dissociation limit from \cite{Krumins:2022}, so it can reproduce also intensity distribution in the experimentally observed bound-free \cstate -\astate transitions. The long range part of the PEC is the same as in Ref.~\cite{Ferber:2013} and it extends the spline point wise form after 12.01 \AA. It congruently converges to the atomic asymptote together with the ground \Xstate state and can reproduce even the highest observed experimental levels ($v_{a}=32$) by taking into account the singlet-triplet mixing. The \emph{ab initio} HFS matrix elements, combined with the empirical \Xstate PEC from \cite{Ferber:2013} and refined \astate PEC in the framework of the simplified coupled-channel deperturbation model, reproduce the experimental term values of both singlet and triplet ground states within 0.003~\cmm accuracies up to their common dissociation limit. So we believe that the present study provides a more accurate estimate of the dissociation energy.

As distinct to the previous studies \cite{Ferber:2009,Ferber:2013}, in this study we used the $R$-dependent HFS coupling functions $A_{0^+-1}(R)$ and $A_{1-1}(R)$ from Ref.~\cite{Oleynichenko:2020} and also from the present\emph{ab initio} calculations of the magnetic hyperfine interaction in \Xstate and \astate states. Due to a more accurate treatment of correlations within the valence and sub-valence shells and a better adaptation of the computational scheme to finite-difference estimation of off-diagonal matrix elements, a better description of $R$-dependencies of HFS parameters than in the previous study \cite{Oleynichenko:2020} is expected. Both diagonal and off-diagonal \emph{ab initio} HFS matrix elements have demonstrated a pronounced $R$-dependence, which respectively supports the observed $v_{a}$-dependence of the HFS.

It turns out that the $R$-dependence of $A_{0^+-1}(R)$ has almost no effect on the fit quality and this could be expected since the coupling between \astate and \Xstate states is significant near the asymptote, where the $A_{0^+-1}(R)$ function approaches the atomic limit. The $R$-dependence of the coupling function $A_{1-1}(R)$, however, allowed for explanation of the observed $v$-dependence of the splitting between the right $G_1=5/2$  and the left $G_1=9/2$ components of the \astate state levels (see Fig.~\ref{split_ave} and also compare Fig.~\ref{AR-comp}a and Fig.~\ref{AR-comp}b). The present calculations were performed with the pure \emph{ab initio} functions. Even better agreement for the splitting (R--L) may be reached if the original \emph{ab initio} $A_{1-1}(R)$ function is scaled as shown in Fig~\ref{split_ave}. Small disagreement for the splittings ($G_{7/2} - G_{9/2}$) and ($G_{5/2} - G_{7/2}$), however, still remains.

The present evaluation of the HFS for the \astate state has demonstrated that the relatively simple CC model presented here (see Fig.~\ref{Hamiltonian_Table}) can be useful when the desired accuracy is of the order of 0.003 \cmm. Its advantage is that the \astate and the \Xstate state levels are modeled by only two potential curves and two radial functions which couple a total of only four channels. The whole HFS of the \astate state is described by the positions of the three $G_1$ components and the nuclear spin of Potassium is neglected.

At higher resolution, the deviations between the CC model and the experiment become visible, namely the model can not reproduce correctly the splitting between the $G_1$ components (see Fig.~\ref{AR-comp}b). The reason for this may be the omitted nuclear spin of Potassium, which leads to additional split of the $G_1$ components and thus to slight change of the overall line profile. In this study we examined this hypothesis by an alternative matrix approach borrowed from Ref.~\cite{Kasahara:1996} and found out that accounting for the K nuclear spin alone brings some improvement but it is not sufficient (Fig.~\ref{A-comp}a). A possible way to resolve the disagreement is to include the matrix elements of the dipole-dipole interaction \cite{Kato:1993} (Fig.~\ref{A-comp}b). The agreement with the experiment becomes satisfactory, however further efforts are needed before a complete agreement is achieved and a physically convincing model is offered. The full HFS deperturbation model should include all significant interactions, line intensities and also the HFS structure of the excited \cstate in order to correctly reproduce the observed line profiles.

% \textbf{Author contributions}:
% 1. Valts Krumins, email valts.krumins@lu.lv, experiment, spectra analysis;
% 2. Maris Tamanis, email maris.tamanis@lu.lv, research methodology, experiment, data analysis, manuscript preparation;
% 3. Ruvin Ferber, email ruvins.ferbers@lu.lv, data analysis, manuscript preparation;
% 4. A. V. Oleynichenko: Methodology, Software, Investigation, Validation, Writing - original draft.
% 5. L. V. Skripnikov: Methodology, Software, Writing - review and editing.
% 6. A. Zaitsevskii: Methodology, Software, Writing - review and editing
% 7. A. Pashov: Conceptualization, Data analysis, Methodology, Software, Writing - original draft, editing
% 8. E. A. Pazyuk: Software, Data analysis, Validation;
% 9. A. V. Stolyarov: Methodology, Writing - review and editing

\section{Acknowledgments}

We are indebted to Ilze Klincare for numerous helpful pieces of advice. Riga team acknowledges the support from the Latvian Council of Science, project No. lzp-2020/2-0215: ``Interatomic potentials of alkali atom pairs in wide range of internuclear distances'' and from the University of Latvia Base Funding No A5-AZ27. AP acknowledges partial support from the Bulgarian Science Fund DN18/12/2017 and from BG05M2OP001-1.002-0019:``Clean technologies for sustainable environment -- waters, waste, energy for circular economy'', financed by the Operational programme ``Science and Education for Smart Growth 2014-2020'', co-financed by the European union through the European structural and investment funds. Moscow team is grateful for the support by the Russian government budget (section 0110), projects No.121031300173-2 and 121031300176-3.

{\small                               %{\footnotesize
\bibliography{ref_KCs}   %>>>> bibliography data in report.bib

\begin{thebibliography}{10}

\bibitem{Ferber:2008}
R.~Ferber, I.~Klincare, O.~Nikolayeva, M.~Tamanis, H.~Kn\"ockel, E.~Tiemann,
  and A.~Pashov.
\newblock The ground electronic state of {KCs} studied by {Fourier} transform
  spectroscopy.
\newblock {\em J. Chem. Phys.}, 128(24):244316, June 2008.

\bibitem{Ferber:2009}
R.~Ferber, I.~Klincare, O.~Nikolayeva, M.~Tamanis, H.~Kn\"ockel, E.~Tiemann,
  and A.~Pashov.
\newblock {$X^{1}\Sigma^{+}$} and {$a^{3}\Sigma^{+}$} states studied by
  {Fourier-transform} spectroscopy.
\newblock {\em Phys. Rev. A}, 80(6):062501, 2009.

\bibitem{Pazyuk:2019}
E.~A. Pazyuk, V.~I. Pupyshev, A.~V. Zaitsevskii, and A.~V. Stolyarov.
\newblock Spectroscopy of diatomic molecules in non-adiabatic approximation.
\newblock {\em Russ. J. Phys. Chem. A}, 93(10):1865--1872, 2019.

\bibitem{Korek:2000}
M.~Korek, A.~R. Allouche, K.~Fakhreddine, and A.~Chaalan.
\newblock Theoretical study of the electronic structure of {LiCs, NaCs, and
  KCs} molecules.
\newblock {\em Can. J. Phys.}, 78(11):977--988, 2000.

\bibitem{Korek:2006}
M.~Korek, Y.~A. Moghrabi, and A.~R. Allouche.
\newblock Theoretical calculation of the excited states of the {KCs} molecule
  including the spin-orbit interaction.
\newblock {\em J. Chem. Phys.}, 124(9):094309, 2006.

\bibitem{Kim:2009}
J.~T. Kim, Y.~Lee, and A.~V. Stolyarov.
\newblock Quasi-relativistic treatment of the low-lying {KCs} states.
\newblock {\em J. Mol. Spectrosc.}, 256(1):57--67, 2009.

\bibitem{Orban:2019}
Orban A., T.~Xie, Vexiau R., O.~Dulieu, and N.~{Bouloufa-Maafa}.
\newblock Hyperfine structure of electronically-excited states of the
  {$^{39}$K$^{133}$Cs} molecule.
\newblock {\em J. Phys. B: At. Mol. Opt. Phys.}, 52(13):135101, 2019.

\bibitem{Kruzins:2010}
A.~Kruzins, I.~Klincare, O.~Nikolayeva, M.~Tamanis, R.~Ferber, E.~A. Pazyuk,
  and A.~V. Stolyarov.
\newblock Fourier-transform spectroscopy and coupled-channels deperturbation
  treatment of the {A$^1\Sigma^+$-b$^3\Pi$} complex of {KCs}.
\newblock {\em Phys. Rev. A}, 81:042509, 2010.

\bibitem{Kruzins:2013}
A.~Kruzins, I.~Klincare, O.~Nikolayeva, M.~Tamanis, R.~Ferber, E.~A. Pazyuk,
  and A.~V. Stolyarov.
\newblock Fourier-transform spectroscopy of {(4)$^1\Sigma^+\rightarrow$}
  {A$^1\Sigma^+$ - b$^3\Pi$}, {A$^1\Sigma^+$ -b$^3\Pi\rightarrow$
  X$^1\Sigma^+$}, and {(1)$^3\Delta_1 \rightarrow $ b$^3\Pi_{0 +/-}$}
  transitions in {KCs} and deperturbation treatment of {A$^1\Sigma^+$} and
  {b$^3\Pi$} states.
\newblock {\em J. Chem. Phys.}, 139:244301, 2013.

\bibitem{Tamanis:2010}
M.~Tamanis, I.~Klincare, A.~Kruzins, O.~Nikolayeva, R.~Ferber, E.~A. Pazyuk,
  and A.~V. Stolyarov.
\newblock Direct excitation of the "dark" {b$^3\Pi$} state predicted by
  deperturbation analysis of the {A$^1\Sigma^+$-b$^3\Pi$} complex in {KCs}.
\newblock {\em Phys. Rev. A}, 82:032506, 2010.

\bibitem{Birzniece:2015}
I.~Birzniece, O.~Nikolayeva, M.~Tamanis, and R.~Ferber.
\newblock Potential construction of the {B(1)$^1\Pi$} state in {KCs} based on
  {Fourier}-transform spectroscopy data.
\newblock {\em J. Quant. Spectrosc. Radiat. Transfer}, 151(0):1 -- 4, 2015.

\bibitem{Szczepkowski:2018}
J.~Szczepkowski, A.~Grochola, P.~Kowalczyk, and W.~Jastrzebski.
\newblock Spectroscopic study of the {(3)$C^1\Sigma^+ \leftarrow X^1\Sigma^+$}
  and {(2)$c^3\Sigma^+ \leftarrow X^1\Sigma^+$} transitions in {KCs} molecule.
\newblock {\em J. Quant. Spectrosc. Radiat. Transfer}, 204:131, 2018.

\bibitem{Kruzins:2021}
A.~Kruzins, V.~Krumins, M.~Tamanis, R.~Ferber, A.~V. Oleynichenko,
  A.~Zaitsevskii, E.~A. Pazyuk, and A.~V. Stolyarov.
\newblock Fourier-transform spectroscopy and relativistic electronic structure
  calculation on the {c$^3\Sigma^+$} state of {KCs}.
\newblock {\em J. Quant. Spectrosc. Radiat. Transf.}, 276:107902, 2021.

\bibitem{Birzniece:2015b}
I.~Birzniece, O.~Nikolayeva, M.~Tamanis, and R.~Ferber.
\newblock Fourier-transform spectroscopy and potential construction of the
  {$(2)^1\Pi$} state in {KCs}.
\newblock {\em J. Chem. Phys.}, 142(13):134309, 2015.

\bibitem{Szczepkowski:2020a}
J.~Szczepkowski, A.~Grochola, P.~Kowalczyk, and W.~Jastrzebski.
\newblock Observation of {D(2)$^1\Pi$ (2)$^3\Pi$ $(2)^3\Sigma^+$} states in
  {KCs} by polarisation labelling spectroscopy technique. modelling of the
  {D(2)$^1\Pi$ (2)$^3\Pi$} system.
\newblock {\em J. Quant. Spectrosc. Radiat. Transf.}, 248:106984, 2020.

\bibitem{Szczepkowski:2013}
J.~Szczepkowski, A.~Grochola, W.~Jastrzebski, and P.~Kowalczyk.
\newblock Study of the {4$^1\Pi$} state in {KCs} molecule by polarisation
  labelling spectroscopy.
\newblock {\em Chem. Phys. Lett.}, 576:10--14, 2013.

\bibitem{Szczepkowski:2015}
J.~Szczepkowski.
\newblock Polarisation labelling spectroscopy of the {$5^1\Pi$} state in {KCs}
  molecule.
\newblock {\em Chem. Phys. Lett.}, 638:78 -- 81, 2015.

\bibitem{Szczepkowski:2020b}
J.~Szczepkowski, A.~Grochola, P.~Kowalczyk, and W.~Jastrzebski.
\newblock Determination of the {C(3)$^1\Sigma^+$} state potential energy curve
  in {KCs} molecule based on polarisation labelling spectroscopy data.
\newblock {\em Spectrochim. Acta A Mol. Biomol. Spectrosc.}, 224:117331, 2020.

\bibitem{Busevica:2011}
Busevica L., Klincare I., Nikolayeva O., Tamanis M., Ferber R., Meshkov~V. V.,
  Pazyuk~E. A., and Stolyarov~A. V.
\newblock Fourier transform spectroscopy and direct potential fit of a
  shelflike state: Application to {E(4)$^1\Sigma^+$} in {KCs}.
\newblock {\em J. Chem. Phys.}, 134:104307, 2011.

\bibitem{Szczepkowski:2012}
J.~Szczepkowski, A.~Grochola, W.~Jastrzebski, and P.~Kowalczyk.
\newblock On the {$4^1\Sigma^+$} state of the {KCs} molecule.
\newblock {\em J. Mol. Spectrosc.}, 276:19--21, 2012.

\bibitem{Szczepkowski:2014}
J.~Szczepkowski, A.~Grochola, W.~Jastrzebski, and P.~Kowalczyk.
\newblock Experimental investigation of the {$6 ^1\Sigma^{+}$} 'shelf' state of
  {KCs}.
\newblock {\em Chem. Phys. Lett.}, 614:36 -- 40, 2014.

\bibitem{Szczepkowski:2019}
J.~Szczepkowski, A.~Grochola, P.~Kowalczyk, W.~Jastrzebski, E.~A. Pazyuk, A.~V.
  Stolyarov, and A.~Pashov.
\newblock The spin-orbit coupling of the {6$^1\Sigma^+$ and 4$^3\Pi$} states in
  {KCs}: observation and deperturbation.
\newblock {\em J. Quant. Spectrosc. Radiat. Transf.}, 239:106650, 2019.

\bibitem{Klincare:2012}
Klincare I., Nikolayeva O., Tamanis M., Ferber R., Pazyuk E.A., and Stolyarov
  A.V.
\newblock Modeling of the {$X^1\Sigma^+$}, {$a^3\Sigma^+ \rightarrow$
  E(4)$^1\Sigma^+ \rightarrow $ X$^1\Sigma^+$(v=0,J=0)} optical cycle for
  ultracold {KCs} molecule production.
\newblock {\em Phys. Rev. A}, 85(6):062520, 2012.

\bibitem{Borsalino:2016}
D.~Borsalino, R.~Vexiau, M.~Aymar, E.~{Luc-Koenig}, O.~Dulieu, and
  N.~{Bouloufa-Maafa}.
\newblock Prospects for the formation of ultracold polar ground state {KCs}
  molecules via an optical process.
\newblock {\em J. Phys. B: At. Mol. Opt. Phys.}, 49:055301, 2016.

\bibitem{Patel:2014}
H.~J. Patel, C.~L. Blackley, S.L. Cornish, and J.~M. Hutson.
\newblock Feshbach resonances, molecular bound states, and prospects of
  ultracold-molecule formation in mixtures of ultracold {K and Cs}.
\newblock {\em Phys. Rev. A}, 90(3):032716, 2014.

\bibitem{Groebner:2016}
M.~Gr\"obner, P.~Weinmann, F.~Meinert, K.~Lauber, E.~Kirilov, and H.-C.
  N\"agerl.
\newblock A new quantum gas apparatus for ultracold mixtures of {K} and {Cs}
  and {KCs} ground-state molecules.
\newblock {\em J. Mod. Opt.}, 63(18):1829--1839, 2016.

\bibitem{Groebner:2017}
M.~Gr\"obner, Ph. Weinmann, E.~Kirilov, H.-Ch. N\"agerl, P.~S. Julienne,
  C.~R.~Le Sueur, and J.~M. Hutson.
\newblock Observation of interspecies feshbach resonances in an ultracold
  {$^{39}$K-$^{133}$Cs} mixture and refinement of interaction potentials.
\newblock {\em Phys. Rev. A}, 95:022715, 2017.

\bibitem{Ferber:2013}
R.~Ferber, O.~Nikolayeva, M.~Tamanis, H.~Kn\"ockel, and E.~Tiemann.
\newblock Long-range coupling of {$X^{1}\Sigma^{+}$} and {$a^{3}\Sigma^{+}$}
  states of the atom pair {K} plus {Cs}.
\newblock {\em Phys. Rev. A}, 88(1):012516, July 2013.

\bibitem{Kasahara:1996}
S.~Kasahara, T.~Ebi, M.~Tanimura, H.~Ikoma, K.~Matsubara, M.~Baba, and
  H.~{Kat\^{o}}.
\newblock High-resolution laser spectroscopy of the {X$^1\Sigma^+$} and
  {(1)$^3\Sigma^+$} states of {$^{23}$Na$^{85}$Rb} molecule.
\newblock {\em J. Chem. Phys}, 105:1341, 1996.

\bibitem{Schwarzer:2021}
M.~Schwarzer and J.~P. Toennies.
\newblock Accurate semiempirical potential energy curves for the
  {a$^3\Sigma^+$} state of {NaCs, KCs, and RbCs}.
\newblock {\em J. Chem. Phys.}, 154:154304, 2021.

\bibitem{Oleynichenko:2020}
A.~V. Oleynichenko, L.~V. Skripnikov, A.~Zaitsevskii, E.~Eliav, and V.~M.
  Shabaev.
\newblock Diagonal and off-diagonal hyperfine structure matrix elements in
  {KCs} within the relativistic fock space coupled cluster theory.
\newblock {\em Chem. Phys. Lett.}, 756:137825, 2020.

\bibitem{Krumins:2022}
V.~Krumins, A.~Kruzins, M.~Tamanis, R.~Ferber, V.~V. Meshkov, E.~A. Pazyuk,
  A.~V. Stolyarov, and A.~Pashov.
\newblock Observation and modeling of bound-free transitions to the
  {$X^1\Sigma^+$} and {$a^3\Sigma^+$} states of {KCs}.
\newblock {\em J. Chem. Phys.}, (accepted), 2022.

\bibitem{Cshfsexp}
E.~Arimondo, M.~Inguscio, and P.~Violino.
\newblock Experimental determinations of the hyperfine structure in the alkali
  atoms.
\newblock {\em Rev. Mod. Phys.}, 49:31, 1977.

\bibitem{FGH}
C.~C. Marston and G.~G. {Balint-Kurti}.
\newblock The {Fourier grid Hamiltonian} method for bound state eigenvalues and
  eigenfunctions.
\newblock {\em J. Chem. Phys.}, 91:3571--3576, 1989.

\bibitem{Rb2depert}
A.~Pashov, P.~Kowalczyk, A.~Grochola, J.~Szczepkowski, and W.~Jastrzebski.
\newblock Coupled-channels analysis of the {5$^1\Sigma^{+}_{\mathrm{u}}$,
  5$^1\Pi_{\mathrm{u}}$, $5^3\Pi_u$, $2^3\Delta_u$} complex of electronic
  states in rubidium dimer.
\newblock {\em J. Quant. Spectrosc. Radiat. Transfer}, 22:225--232, 2018.

\bibitem{Titov:2005}
A.~V. Titov, N.~S. Mosyagin, A.~N. Petrov, and T.~A. Isaev.
\newblock Two-step method for precise calculation of core properties in
  molecules.
\newblock {\em Int. J. Quantum Chem.}, 104(2):223--239, 2005.

\bibitem{Bormotova:2020}
E.~A. Bormotova, A.~V. Stolyarov, L.~V. Skripnikov, and A.~V. Titov.
\newblock Ab initio study of r-dependent behavior of the hyperfine structure
  parameters for the {(1)$^{1,3}\Sigma^+$} states of {LiRb} and {LiCs}.
\newblock {\em Chem. Phys. Lett.}, 760:137998, 2020.

\bibitem{Norman:2018}
P.~Norman, K.~Ruud, and T.~Saue.
\newblock {\em {Principles and Practices of Molecular Properties}}.
\newblock John Wiley {\&} Sons, Ltd, 2018.

\bibitem{Mosyagin:2010}
N.~Mosyagin, A.~Zaitsevskii, and A.~Titov.
\newblock Shape-consistent relativistic effective potentials of small atomic
  cores.
\newblock {\em Int. Rev. At. Mol. Phys.}, 1:63--72, 2010.

\bibitem{RPP-website}
N. S. Mosyagin and A. V. Titov. Generalized relativistic effective core
  potentials. http://www.qchem.pnpi.spb.ru/recp (accessed on 14 January 2022).

\bibitem{Zaitsevskii:2017}
A.~Zaitsevskii, N.~S. Mosyagin, A.~V. Stolyarov, and E.~Eliav.
\newblock {Approximate relativistic coupled-cluster calculations on heavy
  alkali-metal diatomics: application to the spin-orbit-coupled A$^1\Sigma^+$
  and b$^3\Pi$ states of {RbCs} and Cs$_2$}.
\newblock {\em Phys. Rev. A}, 96(2):022516, 2017.

\bibitem{Zaitsevskii:2018}
A.~V. Zaitsevskii, L.~V. Skripnikov, A.~V. Kudrin, A.~V. Oleinichenko,
  E.~Eliav, and A.~V. Stolyarov.
\newblock Electronic transition dipole moments in relativistic coupled-cluster
  theory: the finite-field method.
\newblock {\em Opt. Spectrosc.}, 124(4):451--456, 2018.

\bibitem{Oleynichenko-EXPT}
A.~V. Oleynichenko, A.~Zaitsevskii, and E.~Eliav.
\newblock Towards high performance relativistic electronic structure modelling:
  the {EXP-T} program package.
\newblock In V.~Voevodin and S.~Sobolev, editors, {\em Supercomputing}, volume
  1331, pages 375--386, Cham, 2020. Springer International Publishing.

\bibitem{EXPT-website}
A.~Oleynichenko, A.~Zaitsevskii, and E.~Eliav, 2021.
\newblock {EXP-T}, an extensible code for {F}ock space relativistic coupled
  cluster calculations (see \url{http://www.qchem.pnpi.spb.ru/expt}) (accessed
  on 14 January 2022).

\bibitem{DIRAC19}
DIRAC, a relativistic ab initio electronic structure program, Release DIRAC19
  (2019), written by A. S. P. Gomes, T. Saue, L. Visscher, H. J. Aa. Jensen,
  and R. Bast, with contributions from I. A. Aucar, V. Bakken, K. G. Dyall, S.
  Dubillard, U. Ekstroem, E. Eliav, T. Enevoldsen, E. Fasshauer, T. Fleig, O.
  Fossgaard, L. Halbert, E. D. Hedegaard, T. Helgaker, J. Henriksson, M. Ilias,
  Ch. R. Jacob, S. Knecht, S. Komorovsky, O. Kullie, J. K. Laerdahl, C. V.
  Larsen, Y. S. Lee, H. S. Nataraj, M. K. Nayak, P. Norman, M. Olejniczak, J.
  Olsen, J. M. H. Olsen, Y. C. Park, J. K. Pedersen, M. Pernpointner, R. Di
  Remigio, K. Ruud, P. Salek, B. Schimmelpfennig, B. Senjean, A. Shee, J.
  Sikkema, A. J. Thorvaldsen, J. Thyssen, J. van Stralen, M. L. Vidal, S.
  Villaume, O. Visser, T. Winther, and S. Yamamoto (see
  http://diracprogram.org). (accessed on 14 January 2022).

\bibitem{Saue:2020}
T.~Saue, R.~Bast, A.~S.~P. Gomes, H.~J.~Aa. Jensen, L.~Visscher, I.~A. Aucar,
  R.~{Di Remigio}, K.~G. Dyall, E.~Eliav, E.~Fasshauer, T.~Fleig, L.~Halbert,
  E.~D. Hedegard, B.~Helmich-Paris, M.~Ilias, C.~R. Jacob, S.~Knecht, J.~K.
  Laerdahl, M.~L. Vidal, M.~K. Nayak, M.~Olejniczak, J.~M.~H. Olsen,
  M.~Pernpointner, B.~Senjean, A.~Shee, A.~Sunaga, and J.~N.~P. {van Stralen}.
\newblock The {DIRAC} code for relativistic molecular calculations.
\newblock {\em J. Chem. Phys.}, 152(20):204104, 2020.

\bibitem{Skripnikov:2011}
L.~V. Skripnikov, A.~V. Titov, A.~N. Petrov, N.~S. Mosyagin, and O.~P. Sushkov.
\newblock Enhancement of the electron electric dipole moment in {Eu}$^{2+}$.
\newblock {\em Phys. Rev. A}, 84(2):022505, 2011.

\bibitem{Skripnikov:2015}
L.~V. Skripnikov and A.~V. Titov.
\newblock Theoretical study of {ThF}$^+$ in the search for {T,P}-violation
  effects: Effective state of a {Th} atom in {ThF}$^+$ and {ThO} compounds.
\newblock {\em Phys. Rev. A}, 91(4):042504, 2015.

\bibitem{Stone:2005}
N.~J. Stone.
\newblock Table of nuclear magnetic dipole and electric quadrupole moments.
\newblock {\em At. Data Nucl. Data Tables}, 90(1):75--176, 2005.

\bibitem{ipaasen}
A.~Pashov, W.~{Jastrz\c{e}bski}, and P.~Kowalczyk.
\newblock {Construction of potential curves for diatomic molecular states by
  the IPA method}.
\newblock {\em Comput. Phys. Commun.}, 128:622, 2000.

\bibitem{EPAPS}
Supplementary materials.

\bibitem{Pashov:2005}
{A. Pashov and O. Docenko and M. Tamanis and R. Ferber and H. Kn\"{o}ckel and
  E. Tiemann}.
\newblock {Potential for modelling cold collisions between {Na(3S)} and
  {Rb(5S)} atoms}.
\newblock {\em Phys. Rev. A}, 72:062505, 2005.

\bibitem{Docenko:2006}
O.~Docenko, M.~Tamanis, J.~Zaharova, R.~Ferber, A.~Pashov, H.~Kn\"ockel, and
  E.~Tiemann.
\newblock The coupling of the {X$^1\Sigma^+$} and {a$^3\Sigma^+$} states of the
  atom pair {Na + Cs} and modelling cold collisions.
\newblock {\em J. Phys. B: At. Mol. Opt. Phys.}, 39:S929, 2006.

\bibitem{Staanum:2007}
P.Staanum, A.~Pashov, H.~Kn\"ockel, and E.~Tiemann.
\newblock The {X$^1\Sigma^+$} and {a$^3\Sigma^+$} states of {LiCs} studied by
  {Fourier-transform spectroscopy}.
\newblock {\em Phys. Rev. A}, 75:042513, 2007.

\bibitem{Pashov:2007}
A.~Pashov, O.~Docenko, M.~Tamanis, R.~Ferber, H.~Kn\"ockel, and E.~Tiemann.
\newblock Coupling of the {X$^1\Sigma^+$} and {a$^3\Sigma^+$} states of {KRb}.
\newblock {\em Phys. Rev. A}, 76:022511, 2007.

\bibitem{Kato:1993}
{H. Kato}.
\newblock {Energy levels and Line intensities of Diatiomic molecules.
  Application to Alkali Metal Molecules.}
\newblock {\em Bull. Chem. Soc. Jpn}, 66:3203, 1993.

\end{thebibliography}
\bibliographystyle{unsrt}  %apsrev}   %>>>> makes bibtex use spiebib.bst

\end{document}